\magnification=1200
\def\qed{\unskip\kern 6pt\penalty 500\raise -2pt\hbox
{\vrule\vbox to 10pt{\hrule width 4pt\vfill\hrule}\vrule}}
\centerline{NONEQUILIBRIUM STATISTICAL MECHANICS AND ENTROPY PRODUCTION}
\centerline{IN A CLASSICAL INFINITE SYSTEM OF ROTATORS.}
\bigskip
\centerline{by David Ruelle\footnote{*}{Mathematics Dept., Rutgers University, and IHES.  91440 Bures sur Yvette, France.\break $<$ruelle@ihes.fr$>$}.}
\bigskip\bigskip\noindent
	{\leftskip=2cm\rightskip=2cm\sl Abstract.  We analyze the dynamics of a simple but nontrivial classical Hamiltonian system of infinitely many coupled rotators.  We assume that this infinite system is driven out of thermal equilibrium either because energy is injected by an external force (Case I) , or because heat flows between two thermostats at different temperatures (Case II).  We discuss several possible  definitions of the entropy production associated with a finite or infinite region, or with a partition of the system into a finite number of pieces.  We show that these definitions satisfy the expected bounds in terms of thermostat temperatures and energy flow.\par}
\vfill\eject
\noindent
{\bf 0 Introduction.}
\medskip
	In the present paper, we study certain classical Hamiltonian systems consisting of an infinite number of coupled degrees of freedom (rotators or ``little wheels'').  For a system in the class considered, the time evolution $(f^t)$ is well defined, and given by the limit (in some sense) of the Hamiltonian time evolution for finite subsystems.  [Note that other infinite systems, like gases of interacting particles, would be much more difficult to control].  A probability measure on the phase space of the infinite system is called a {\it state}, and it has a well-defined time evolution.  We introduce a family of initial states called $\Gamma$-states (they are Gibbs states of some sort).  Some of these $\Gamma$-states describe a situation where parts of our infinite system (thermostats) are at given temperatures.  For a $\Gamma$-state $\ell$, the time-evolved state $f^t\ell$ gives a finite Gibbs entropy $S^t(X)$ to each finite subsystem $X$ of the infinite system $L$.  If $X$ is infinite (but has finite interaction with the rest of the system) the difference $\Delta S^t(X)=\lim_{Y\to\infty}(S^t(X\cap Y)-S^0(X\cap Y))$ still makes sense.
\medskip
	The bulk of the paper is dedicated to a discussion of the (nontrivial) dynamics of our infinite system of rotators.  Understanding the dynamics of the system is a necessary prerequisite to analyzing its nonequilibrium statistical mechanics.  We shall in fact examine a specific nonequilibrium problem: is it possible to define a local rate of entropy production (associated with a finite region $X$) in a nontrivial manner?  This possibility has been suggested by Denis Evans and coworkers [16].  We examine their proposal and some alternatives, but obtain only partial results.  Because of the obvious physical interest of the problem, we now give some details.
\medskip
	By time-averaging $f^t\ell$ or $d\Delta S^t(X)/dt$ (over a suitable sequence of intervals $[0,T]\to\infty$) we may define a nonequilibrium steady state 
$$	\rho=\lim_{T\to\infty}{1\over T}\int_0^Tdt\,f^t\ell      $$
and an average rate of entropy growth
$$	\sigma(X)=\lim_{T\to\infty}{1\over T}\Delta S^T(X)      $$
(we do not know that $\sigma(X)$ is uniquely determined by $\rho$ and $X$).
\medskip
	We ask if an entropy production rate $e(X)$ can be meaningfully associated with a finite set $X\subset L$.  For definiteness we shall think of two physical situations.  In Case I there is a finite set $X_0$ such that an external force acts on $X_0$, and the initial state $\ell$ restricted to $L\backslash X_0$ corresponds to thermal equilibrium at temperature $\beta^{-1}$.  In Case II we have $L=X_0\sqcup L_1\sqcup L_2$ where $X_0$ is finite, $L_1$ and $L_2$ are infinite and $\ell$ restricted to $L_i$ corresponds to thermal equilibrium at temperature $\beta_i^{-1}$ (with $\beta_1^{-1}<\beta_2^{-1}$).  There is a thermodynamic formula for the global rate of entropy production:
$$	e_\Theta=\beta\times\hbox{energy flux to thermostat}\qquad\hbox{(Case I)}      $$
$$	e_\Theta=(\beta_1-\beta_2)
	\times\hbox{energy flux to thermostat 1}\qquad\hbox{(Case II)}      $$
[Note that Case I resembles Case II, where thermostat 2 is replaced by the external force, and ascribed an infinite temperature ($\beta_2=0$)].  The question is how to define a local rate of entropy production $e(X)\ge0$ such that $\sup_{X\,{\rm finite}}e(X)=e_\Theta$.
\medskip
	The original proposal by Evans and coworkers\footnote{*}{Actually, the ideas presented in [16] are formulated for Case I, and for a system thermostatted at the boundary rather than an actually infinite system.  While the two idealizations are technically quite different, they are expected to give the same results in cases of physical interest.} is to take, for $X$ finite, 
$$	e(X)=-\sigma(X)      $$
This is shown to be the average rate of volume contraction in the phase space $[X]$ of the subsystem $X$ due to the fluctuating forces to which it is subjected by the complementary subsystem $L\backslash X$.
\medskip
	Another idea is to replace the entropy $S(X)$ by the conditional entropy given formally by $\check S(X)=S(L)-S(L\backslash X)$.  The corresponding rate of entropy production is 
$$	\check e(X)=\sigma(L\backslash X)      $$
\indent
	We shall make the important physical assumption that the expectation value of the energy for each finite system $X$ has a bound independent of time\footnote{**}{Note that in Case I, if the system has dimension $\le2$, the external force may cause an infinite accumulation of energy in a finite region.  Our assumption that the nonequilibrium steady state $\rho$ gives a finite expectation to the energy of finite subsystems is thus invalid, and so is our analysis.}.  It follows that $\check e(X)$ is finite, and one has
$$	0\le e(X)\le\check e(X)      $$
\medskip
	Instead of using a finite set $X$ one may base a definition of entropy production rate on a finite partition ${\cal A}=(X_0,X_1,\ldots,X_n)$ of $L$, with finite boundary (this will be made precise later).  We define
$$	e({\cal A})=\sum_{j=0}^n\sigma(X_j)\qquad,
	\qquad\check e({\cal A})=\sum_{j:X_j\,{\rm infinite}}\sigma(X_j)      $$
In particular, in Case II, for $X$ finite $\supset X_0$, we have
$$	\check e(X)=\check e((X,L\backslash X))
	\le\check e((X,L_1\backslash X,L_2\backslash X))      $$
and the right-hand side $e((X,L_1\backslash X,L_2\backslash X))$ seems a rather  natural definition of entropy production rate.
\medskip
	We shall later study further properties of the entropy production rates defined above, but we note here that they are all bounded by the thermodynamic expression $e_\Theta$.  The problem is to prove that they depend effectively on $X$ or ${\cal A}$, and are not identically equal to $0$ or $e_\Theta$.
\medskip
	We now recall some earlier work to put the problem of defining a local entropy production rate in perspective.
\medskip
	In earlier studies of quantum spin systems [15], [11], the global entropy production (for Case II) was defined by the thermodynamic relation 
$$	e_\Theta=(\beta_1-\beta_2)\times\hbox{energy flux to thermostat 1}      $$
but the quantities $e(X)$, $\check e(X)$ were not introduced because they would automatically vanish.  This is because, for quantum spin systems we have $|\check S^t(X)|<S^t(X)$ (see [3] Proposition 6.2.28(b)); for classical rotators by contrast, the entropy is not bounded below.
\medskip
	The statistical mechanics of classical systems outside of equilibrium can be studied in models with nongradient forces and a ``deterministic thermostat'' [7], [10].  Such a nonhamiltonian system corresponds in effect to a rather general time evolution $(f^t)$ defined by a vector field ${\cal X}$ on a finite dimensional manifold $M$.  In general, no absolutely continuous invariant measure ({\it i.e.}, ``phase space volume'' $m$) on $M$ is preserved by the time evolution, but one may assume that there is a natural (singular) measure $\rho$ describing a nonequilibrium steady state.  One can argue that the average phase space volume contraction $\int\rho(dx)(-{\rm div}_m{\cal X})(x)$ is the rate of entropy production by the system.  This identification (for which see Andrei [1]) has been used in particular by Evans, Cohen, and Morriss [6], and by Gallavotti and Cohen [9] in the study of fluctuations of the entropy production.  See also the work of Posch and Hoover [13], Gallavotti [8].
\medskip
	Note now that if we introduce a nongradient force $\xi(q)$ in the Hamiltonian equations of motion, the volume $dp\,dq$ is preserved, but energy conservation is lost and this is why a thermostat is needed.  In the case of a deterministic thermostat, the phase space contraction is caused by the thermostat (as one can check in the example of the {\it isokinetic thermostat} corresponding to an added ``force'' $-\alpha(p,q)p$, where $\alpha(p,q)=p\cdot\xi(q)/p\cdot p$).  In the lab however the thermostat is of a different nature: it is typically a large system (reservoir) with which the small system of interest can exchange heat, and it is not clear at first how to define entropy production.  In particular, a nonequilibrium steady state for the infinite system $L$ may well have absolutely continuous projection on the phase space of the small system $X$ [4], [5], [2], which contradicts $e(X)>0$ but may allow $\check e(X)>0$.\medskip
	Finally, to indicate the difficulty of the problems considered here, and in particular of proving $\check e(X)>0$, consider Case II in dimension $\le2$.  There (as indicated by the macroscopic continuous limit), $f^t\ell$ presumably tends to an {\it equilibrium} state $\rho$ and the entropy production $\check e(X)$ vanishes for all $X$.
\bigskip\noindent
{\bf Acknowledgments.}
\medskip
	During the lengthy elaboration of the present paper, I have benefitted from useful correspondence with D. Evans, and many discussions with J.-P. Eckmann, G. Gallavotti, J.L. Lebowitz, H. Moriya, H. Posch, and L.-S. Young.
\bigskip\noindent
{\bf 1 Description of the model.}
\medskip
	Our system will be an infinite collection of rotators labelled by $x\in L$, each with Hamiltonian $H_x(p_x,q_x)=p_x^2/2+V_x(q_x)$, where $p_x\in{\bf R}$, $q_x\in{\bf T}$. [This is for simplicity; it would probably be easy to replace the rotators by more complicated systems].  We let $\Gamma$ be a set of unordered pairs $\{x,y\}$ of points in $L$, {\it i.e.}, $\Gamma$ is a graph with vertex set $L$, and we define a formal Hamiltonian for the infinite system of little wheels: 
$$   \sum_{x\in L}H_x(p_x,q_x)+\sum_{\{x,y\}\in\Gamma}W_{\{x,y\}}(q_x,q_y)   $$
The functions $V_x$, $W_{\{x,y\}}$ are assumed to be smooth.
\medskip
	For $X\subset L$, let $\Gamma_X=\{\{x,y\}\in\Gamma:x,y\in X\}$ and, when $X$ is finite, write 
$$	H_X(p_X,q_X)=\sum_{x\in X}H_x(p_x,q_x)
	+\sum_{\{x,y\}\in\Gamma_X}W_{\{x,y\}}(q_x,q_y)      $$
where $p_X=(p_x)_{x\in X}\in{\bf R}^X$, $q_X=(q_x)_{x\in X}\in{\bf T}^X$.  We shall also make use of a constant external force\footnote{*}{$F$ is taken constant for simplicity.  More generally one could consider the case of a smooth function $F(q_{X_0},\phi^t\alpha)$ of $q_{X_0}$ and $\phi^t\alpha$ with values in ${\bf R}^{X_0}$, where $(\phi^t)$ is a smooth dynamical system on a compact manifold ${\cal A}$, and $\alpha$ is distributed according to some prescribed $(\phi^t)$-ergodic measure on ${\cal A}$.}  $F\in{\bf R}^{X_0}$ acting on a finite set $X_0$.
\medskip
	For finite $X$, a time evolution $(f_X^t)$ on ${\bf R}^X\times{\bf T}^X$ is defined by
$$	{d\over dt}\pmatrix{p_X\cr q_X\cr}
=\pmatrix{F_X-\partial_{q_X}H_X(p_X,q_X)\cr p_X\cr}      $$
where the term $F_X$ is the component of $F$ in ${\bf R}^X$, and is present only in case I.  We have thus
$$	f_X^t(p_X(0),q_X(0))=(p_X(t),q_X(t))      $$
\indent
	We shall suppose that $\Gamma$ is connected and, for $x,y\in L$, define
$$	d(x,y)=\min\{k:\exists x_0,\ldots,x_k\in L\hbox{ with }x_0=x,x_k=y
	\hbox{ and }\{x_{j-1},x_j\}\in\Gamma\hbox{ for }j=1,\ldots k\}      $$ 
We write then $B_x^k=\{y:d(x,y)\le k\}$.
\medskip
	{\bf 1.1 Assumption} (finite dimensionality).

	{\sl There is a polynomial $P(k)$ such that for all $x\in L$ and $k\ge0$ 
$$	|B_x^k|\le P(k)      $$}
[We may take $P(k)=1+ak^b$ for some $a,b>0$; $\Gamma$ is thus assumed to have order $\le a$, and ``dimension'' $\le b$]. 
\medskip
	Note that by compactness of ${\bf T}$ and the assumed smoothness of $V_x$, $W_{\{x,y\}}$, every ``force'' term ({\it i.e.}, each component $F_x$ of $F$ for $x\in X_0$, each $\partial_{q_x}V_x$, each $\partial_{q_x}W_{\{x,y\}}$, $\partial_{q_y}W_{\{x,y\}}$) is bounded, and has bounded derivatives with respect to its arguments $q_z$.
\medskip
	{\bf 1.2 Assumption} (uniform boundedness).

	{\sl The force terms $\partial_{q_x}V_x$, $\partial_{q_x}W_{\{x,y\}}$, $\partial_{q_y}W_{\{x,y\}}$ and their $q_z$-derivatives (up to any finite order) are bounded uniformly in $x,y\in L$.}
\medskip
	{\bf 1.3 Lemma} (uniform boundedness of forces).

	{\sl The forces $F_x-\partial_{q_x}H_X(p_X,q_X)$ or $-\partial_{q_x}H_X(p_X,q_X)$ and their $q_z$-derivatives have modulus bounded respectively by constants $K$, $K'$.  [We shall also denote by $\bar K$ a constant $\ge2K,P(1)K'$,1].}
\medskip
	This follows from Assumptions 1.1 and 1.2 [only the bounded order of $\Gamma$ is used from Assumption 1.1].\qed
\bigskip\noindent
{\bf 2 Time evolution of infinite systems.}
\medskip
	For $X\subset L$, we shall from now on write $[X]=({\bf R}\times{\bf T})^X$.  We note the following facts which follow from Lemma 1.3.
\medskip
	(i) For $X$ finite and $\xi\in[X]$, if $f_X^t\xi=(p_x(t),q_x(t))_{x\in X}$ we have the estimate
$$	|p_x(t)-p_x(0)|\le K|t|\quad{\rm when}\quad x\in X      $$
\indent
	(ii) For $\tilde X$ finite and $\tilde\xi\in[\tilde X]$ let also $f_{\tilde X}^t\tilde\xi=(\tilde p_x(t),\tilde q_x(t))_{x\in\tilde X}$.  Then, if $p_x(0)=\tilde p_x(0)$, $q_x(0)=\tilde q_x(0)$ for some $x\in X\cap\tilde X$ we have
$$	|p_x(t)-\tilde p_x(t)|\le2K|t|      $$
$$	|q_x(t)-\tilde q_x(t)|\le K|t|^2      $$
and since $|q_x(t)-\tilde q_x(t)|\le1\le\bar K$, we also have 
$$	|q_x(t)-\tilde q_x(t)|\le[K|t|^2.\bar K]^{1\over2}\le\bar K|t|      $$
so that
$$	\max(|p_x(t)-\tilde p_x(t)|,|q_x(t)-\tilde q_x(t)|)\le\bar K|t|      $$
\indent
	(iii) Let $k\ge0$ and $X\supset B_x^k$, $\tilde X\supset B_x^k$.  Then, with the notation of (ii), if 
$$  (\forall y\in B_x^k)\qquad p_y(0)
	=\tilde p_y(0)\quad{\rm and}\quad q_y(0)=\tilde q_y(0)      $$
we have
$$	\max(|p_x(t)-\tilde p_x(t)|,|q_x(t)-\tilde q_x(t)|)
	\le{(\bar K|t|)^{k+1}\over(k+1)!}      $$
indeed, by the equation of motion and induction on $k$ we have 
$$	|{d\over dt}(f_X^t\xi-f_X^t\tilde\xi)|
	\le \bar K{(\bar K|t|)^k\over k!}      $$
and the desired result follows by integration.
\medskip
	{\bf 2.1 Lemma} (a priori estimates).
\medskip
	{\sl For finite $X,\tilde X\subset L$, let $\xi\in[X]$, $\tilde\xi\in[\tilde X]$, and $f_X^t\xi=(\xi_x(t))_{x\in X}=(p_x(t),q_x(t))$, $f_{\tilde X}^t\tilde\xi=(\tilde\xi_x(t))_{x\in\tilde X}=(\tilde p_x(t),\tilde q_x(t))$.  With this notation,
	
\noindent
(a) $\qquad|p_x(t)-p_x(0)|\le K|t|$, hence $|p_x(t)|\le|p_x(0)|+K|t|$

\noindent
(b) if $k>0$, and $B_x^{k-1}\subset X\cup\tilde X$, and $\xi_y(0)=\tilde\xi_y(0)$ for all $y\in B_x^{k-1}$, then
$$|\xi_x(t)-\tilde\xi_x(t)|=\max[|p_x(t)-\tilde p_x(t)|,|q_x(t)-\tilde q_x(t)|]
	\le{(\bar K|t|)^k\over k!}      $$}
\indent
	This follows from (i) and (iii) above [(b) is a rather rough estimate, but sufficient for our purposes].\qed
\medskip
	{\bf 2.2 Proposition} (time evolution).
\medskip
	{\sl Let $(p_x(0),q_x(0))_{x\in L}\in[L]$ be given.  For finite $X\subset L$, write $(p_x^X(t),q_x^X(t))_{x\in X}=f_X^t(p_x^X(0),q_x^X(0))_{x\in X}$.  Then for each $x\in L$ the limits
$$	\lim_{d(x,L\backslash X)\to\infty}p_x^X(t)=p_x(t)\qquad,\qquad
	\lim_{d(x,L\backslash X)\to\infty}q_x^X(t)=q_x(t)      $$
exist, and $(p_x(t),q_x(t))_{x\in L}$ is the unique solution of the infinite system evolution equation with initial condition $(p_x(0),q_x(0))_{x\in L}$.  We write $(p_x(t),q_x(t))_{x\in L}=f^t(p_x(0),q_x(0))_{x\in L}$.}
\medskip
	The existence of the limit follows from Lemma 2.1.  Writing the infinite system evolution equation is left to the reader, as well as checking that $(p_x(t),q_x(t))_{x\in L}$ is the unique solution.\qed
\medskip
	{\bf 2.3 Remarks.}
\medskip
	The limits in Proposition 2.2 are faster than $\exp(-k\,d(x,L\backslash X))$ for any $k>0$, independently of $(p_x(0),q_x(0))_{x\in L}$, and uniformly for $t$ in any compact interval $[-T,T]$.
\medskip
	Existence and uniqueness theorems are known in more difficult situations; see for instance [12].
\medskip
	The proof of Proposition 2.2 does not use the finite dimensionality of $\Gamma$, only its finite order.
\medskip
	{\bf 2.4 Notation.}
\medskip
	In principle we use the notation $(p_x^X(t),q_x^X(t))_{x\in X}$ for the finite system time evolution $(f_X^t)$, and $(p_x(t),q_x(t))_{x\in L}$ for the infinite system evolution $(f^t)$, but it will often be convenient to drop the superscript $X$.
\medskip
	It is useful to compactify the momentum space ${\bf R}$ to a circle $\dot{\bf R}$ by addition of a point at infinity for each $x\in L$, and write $[\dot X]=(\dot{\bf R}\times{\bf T})^X$ for $X\subset L$.  The phase space of our infinite system is then $[L]=({\bf R}\times{\bf T})^L\subset(\dot{\bf R}\times{\bf T})^L=[\dot L]$.  We shall use the product topologies on $[L]=({\bf R}\times{\bf T})^L$ and $[\dot L]=(\dot{\bf R}\times{\bf T})^L$; therefore $[\dot L]$ is compact and $[L]$ has the topology it inherits as subset of $[\dot L]$.  If $U\subset L$ we denote by $\pi_U$ the projection
$$	\pi_U:[\dot L]=[\dot U]\times[\dot{L\backslash U}]\to[\dot U]      $$
\indent
	{\bf 2.5 Proposition} (continuity of $f^t$).
\medskip
	{\sl The map $(\xi,t)\mapsto f^t\xi$ is continuous $[L]\times{\bf R}\to[L]$ and, for each $t$, $f^t:[L]\to[L]$ is a homeomorphism.}
\medskip
	To prove the continuity of $(\xi,t)\mapsto f^t\xi$, it suffices to prove the continuity of $(\xi,t)\mapsto(p_x(t),q_x(t))$ for each $x\in L$, and this results from the uniformity of the limits in Proposition 2.2 (see Remark 2.3).  By uniqueness of $f^t$, the map $f^{-t}$ is the inverse of $f^t$ and, since $f^{-t}:[L]\to[L]$ is continuous, $f^t$ is a homeomorphism.\qed
\medskip
	{\bf 2.6 Proposition} (smoothness of $f^t$).
\medskip
	{\sl Let $X\subset Y$, $X$ finite and $Y$ finite or $=L$.  For $\xi\in[X]$, $\eta\in[Y\backslash X]$, write $f_Y^t(\xi,\eta)=(p_x(t),q_x(t))_{x\in Y}$.  Then, for fixed $\eta$ and each $x\in Y$, the map $(\xi,t)\mapsto(p_x(t),q_x(t))$ is smooth $[X]\times{\bf R}\to{\bf R}\times{\bf T}$.}  
\medskip
	This results from the bounds on the derivatives (uniform in $Y$) obtained in Proposition 2.7 below.
\medskip
	{\bf 2.7 Proposition} (estimate of derivatives).
\medskip
	{\sl Let $Y\subset L$, $Y$ finite or $=L$, and
$$	f_Y^t(\eta_x(0))_{x\in Y}=(\eta_x(t))_{x\in Y}=(p_x(t),q_x(t))      $$
Write
$$      r_x^{(i,j)}(t;x_1,\ldots,x_j)={\partial^i\over\partial t^i}
        {\partial\over\partial\eta_{x_1(0)}}\ldots
        {\partial\over\partial\eta_{x_j(0)}}\eta_x(t)      $$
where $x_1,\ldots,x_j$ need not be all distinct; then
$$      |r_x^{(i,j)}(t;x_1,\ldots,x_j)|\le{\cal P}_{ij}((|p_y(0)|+K|t|))     $$
where ${\cal P}_{ij}((p_y))$ is a polynomial of degree $\le i+1$ (the degree is $1$ if $(i,j)=(0,0)$, $\le i$ otherwise) in the $p_y$ with $y\in B_x^i$.
\medskip
        Let $\sigma=\sigma(x,x_1,\ldots,x_j)$ denote the smallest number of edges of a connected subgraph of $\Gamma$ having $x,x_1,\ldots,x_j$ among its vertices.  Then the coefficients of ${\cal P}_{ij}$ are positive and
$$      \le M_{ij}e^{j\bar K|t|}{(L_j|t|)^\tau\over\tau!}      $$
with suitable $L_j,M_{ij}>0$, for all $\tau$ such that $0\le\tau\le[\sigma-i]_+$ where we have written $[\sigma-i]_+=\max(0,\sigma-i)$.}
\medskip
        The proof is given in Appendix A.1.
\medskip
        {\bf 2.8 Proposition} (estimate of differences).
\medskip
        {\sl We use the notation of Proposition 2.7.  Let $(\eta_x(0)),(\tilde\eta_x(0))\in[Y]$ and define $\tilde r_x^{(i,j)}$ as $r_x^{(i,j)}$with $\eta$ replaced by $\tilde\eta$.  For finite $X\subset L$ we assume $\eta_y(0)=\tilde\eta_y(0)$ when $y\notin X$, and write
$$      \Delta r_x^{(i,j)}(t;x_1,\ldots,x_j;X)
=r_x^{(i,j)}(t;x_1,\ldots,x_j;X)-\tilde r_x^{(i,j)}(t;x_1,\ldots,x_j;X)      $$
If $d(x,X)>i$, we have
$$      |\Delta r_x^{(i,j)}(t;x_1,\ldots,x_j;X)|\le{\cal Q}((|p_y(0)|+K|t|))      $$
where ${\cal Q}((p_y))$ is a polynomial of degree $\le i+1$ (the degree is $1$ if $(i,j)=(0,0)$, $\le i$ otherwise) in the $p_y$ with $y\in B_x^i$.
\medskip
        Let $\sigma=\sigma(x,x_1,\ldots,x_j;X)$ denote the smallest number of edges of a subgraph of $\Gamma$ (not necessarily connected) connecting each point $x,x_1,\ldots,x_j$ to some point of $X$.  Then the coefficients of ${\cal Q}_{ij}$ are positive and
$$      \le M_{ij}e^{j\bar K|t|}{(L_j|t|)^\tau\over\tau!}      $$
for all $\tau$ such that $0\le\tau\le\sigma-i$}
\medskip
        The proof is given in Appendix A.2.
\medskip
        {\bf 2.9 Remarks}
\medskip
        Proposition 2.7, 2.8 will be used in the proof of Theorem 4.5 below.  In view of these applications the following facts should be noted.
\medskip
        (a) The condition $d(x,X)>i$ in Proposition 2.8 is not a serious limitation because, for the finitely many values of $x$ such that $d(x,X)\le i$, one can estimate $\Delta r^{(i,j)}$ by Proposition 2.7 applied to $r^{(i,j)}$ and $\tilde r^{(i,j)}$.
\medskip
        (b) Write $\sigma=\sigma(y,y_1,\ldots,y_j)$ and let $y$ be fixed, then
$$        \sum_{y_1,\ldots,y_j}{1\over(\sigma-i)!}<\infty        $$
Indeed, we have $|y_k-y|\le\sigma$, hence
$$      \sigma-i\ge\sum_{k=1}^j{|y_k-y|-i\over j}      $$
so that $1/(\sigma-i)!$ decreases faster than exponentially with respect to $r_1=|y_1-y|,\ldots,r_j=|y_j-y|$, while $|B_y^{r_1}|\cdots|B_y^{r_j}|$ is polynomially bounded.
\bigskip\noindent
{\bf 3 Time evolution for probability measures.}
\medskip
	Consider any probability measure $\ell$ on $[\dot L]=(\dot{\bf R}\times{\bf T})^L$ carried by $[L]=({\bf R}\times{\bf T})^L$ ({\it i.e.}, $\ell$ gives zero measure to the points at infinity).  We can find constants $\kappa_{nx}>0$ such that, if we write
$$   B_n=\{(p_x,q_x)_{x\in L}:|p_x|\le\kappa_{nx}\hbox{ for all }x\in L\}   $$
we have $\ell(B_n)>1-1/n$.  We may thus write $\lim_{n\to\infty}||\ell-\ell_n||=0$ where the measure $\ell_n$ has support in the compact set $B_n\subset[L]$, and $(t,\xi)\to f^t\xi$ is continuous on ${\bf R}\times B_n$.  We define then
$$	f^t\ell=\lim_{n\to\infty}f^t\ell_n\qquad\hbox{(norm limit)}      $$
Notice also that $f^t\ell_n$ has support in the compact set $B'_n$ defined like $B_n$ with $\kappa_{nx}$ replaced by $\kappa'_{nx}=\kappa_{nx}+K|t|$ (see Lemma 2.1(a)).  Therefore $f^t\ell$ is again carried by $[L]$.
\medskip
	{\bf 3.1 Proposition} (continuity of time evolution).
\medskip
	{\sl If the probability measure $\ell$ on $[\dot L]$ is carried by $[L]$, then the probability measure $f^t\ell$ is well defined, carried by $[L]$, and $t\mapsto f^t\ell$ is continuous ${\bf R}\to$ measures on $[\dot L]$ with the vague topology.}
\medskip
	For any continuous function $A:[\dot L]\to{\bf R}$ we have $(f^t\ell_n)(A)=\ell_n(A\circ f^t)$, where $A\circ f^t$ restricted to $B_n$ depends continuously on $t$ with respect to the uniform norm on ${\cal C}(B_n\to{\bf R})$.  Therefore $(f^t\ell_n)(A)$ is a continuous function of $t$, and so is its uniform limit $t\mapsto f^t\ell(A)$.  This shows that $t\mapsto f^t\ell$ is continuous with respect to the $w^*$ (=vague) topology of measures on $[\dot L]$, concluding the proof.\qed
\medskip
	Let $\ell_X$ be a probability measure on $[\dot X]$ for finite $X\subset L$.  We write $X\to\infty$ when, for every finite $U\subset L$, eventually $X\supset U$.  Suppose that for every finite $U$ and $A\in{\cal C}([\dot U]\to{\bf R})$ the limit
$$	\lim_{X\to\infty}\ell_X(A\circ\pi_{UX})      $$
exists, where $\pi_{UX}$ is the projection $[\dot X]=[\dot{X\backslash U}]\times[\dot U]\to[\dot U]$.  This limit is then of the form $\ell(A\circ\pi_U)$ where $\ell$ is a uniquely defined probability measure on $[\dot L]$ which we call the {\it thermodynamic limit} of the $\ell_X$:
$$	\ell={\theta\lim}_{X\to\infty}\ell_X      $$
This means that 
$$	\pi_U\ell={w^*\lim}_{X\to\infty}\pi_{UX}\ell_X      $$
or (modulo the identifications $A\to A\circ\pi_{UX}$, $A\to A\circ\pi_U$) $\ell$ is the limit of the $\ell_X$ on 
$$    \cup_{U\,\hbox{finite}}\,{\cal C}([\dot U]\mapsto{\bf R})\circ\pi_U    $$
which is dense in ${\cal C}([\dot L]\to{\bf R})$.  In particular, if $\ell$ is any probability measure on $[\dot L]$, we have
$$	\ell={\theta\lim}_{X\to\infty}\pi_X\ell      $$
We shall later also consider thermodynamic limits associated with a sequence $X_n\to\infty$, writing $\ell=\theta\lim_{n\to\infty}\ell_{X_n}$ if $\pi_U\ell=w^*\lim_{n\to\infty}\pi_{UX_n}\ell_{X_n}$ for all finite $U\subset L$.
\medskip
	{\bf 3.2 Proposition} (time evolution of thermodynamic limits).
\medskip
	{\sl Suppose that
$$	{\theta\lim}_{X\to\infty}\ell_X=\ell      $$
where $\ell_X,\ell$ are probability measures carried by $[X],[L]$ respectively.  Then 
$$	{\theta\lim}_{X\to\infty}f_X^t\ell_X=f^t\ell      $$
uniformly for $t\in[-T,T]$.}
\medskip
	We have to prove that, for every finite $U\subset L$, and $A\in{\cal C}([\dot U]\to{\bf R})$,
$$   \lim_{X\to\infty}(f_X^t\ell_X)(A\circ\pi_{UX})=(f^t\ell)(A\circ\pi_U)   $$
We may (and shall) assume that $|A|\le1$.  Given $\epsilon>0$, we know that
$$	||A\circ\pi_U\circ f^t
	-A\circ\pi_{UX}\circ f_X^t\circ\pi_X||<{\epsilon\over2}      $$
for sufficiently large X, say $X\supset V$ for suitable $V\supset U$, for all $t\in[-T,T]$.  Under these conditions we have thus
$$      ||A\circ\pi_U\circ f^t-A\circ\pi_{UV}\circ f_V^t\circ\pi_V||
        <\epsilon/2      $$
and
$$	||A\circ\pi_{UX}\circ f_X^t\circ\pi_X
	-A\circ\pi_{UV}\circ f_V^t\circ\pi_V||<\epsilon      $$
which we shall use below in the form
$$	||A\circ\pi_{UX}\circ f_X^t
	-A\circ\pi_{UV}\circ f_V^t\circ\pi_{VX}||<\epsilon      $$
\indent
	Take now a function $\Phi\in{\cal C}([V]\to{\bf R})$ with compact support and $|\Phi|\le1$, such that
$$	||\ell-(\Phi\circ\pi_V)\ell||<\epsilon      $$
Using the notation $a\buildrel\epsilon\over\sim b$ to mean $|a-b|<\epsilon$, we have
$$	(f^t\ell)(A\circ\pi_U)\buildrel\epsilon\over\sim
	(f^t((\Phi\circ\pi_V)\ell))(A\circ\pi_U)
	=\ell((\Phi\circ\pi_V)(A\circ\pi_U\circ f^t))      $$
$$	\buildrel\epsilon\over\sim
	\ell((\Phi\circ\pi_V)(A\circ\pi_{UV}\circ f_V^t\circ\pi_V))
	=\ell((\Phi(A\circ\pi_{UV}\circ f_V^t))\circ\pi_V)
	=\ell(\Psi_t\circ\pi_V)      $$
where the function $\Psi_t=\Phi(A\circ\pi_{UV}\circ f_V^t):[V]\to{\bf R}$ is continuous with compact support, hence extends to a continuous function on $[\dot V]$.  By assumption we have 
$$	|\ell(\Psi_t\circ\pi_V)-\ell_X(\Psi_t\circ\pi_{VX})|<\epsilon      $$
for sufficiently large $X$, uniformly with respect to $t\in[-T,T]$ (this is because $t\to\Psi_t$ is continuous with respect to the uniform norm of ${\cal C}([\dot V]\to{\bf R})$).  We may thus take $W\supset V$ such that, if $X\supset W$ and $t\in[-T,T]$,
$$	\ell(\Psi_t\circ\pi_V)\buildrel\epsilon\over\sim
	\ell_X(\Psi_t\circ\pi_{VX})
	=\ell_X((\Phi(A\circ\pi_{UV}\circ f_V^t))\circ\pi_{VX})      $$
$$	\buildrel\epsilon\over\sim
	\ell_X((\Phi\circ\pi_{VX})(A\circ\pi_{UX}\circ f_X^t))
	=(f_X^t((\Phi\circ\pi_{VX})\ell_X))(A\circ\pi_{UX})      $$
We have thus
$$	|(f^t\ell)(A\circ\pi_U)
	-(f_X^t((\Phi\circ\pi_{VX})\ell_X))(A\circ\pi_{UX})|<4\epsilon      $$
when $X\supset W$, $t\in[-T,T]$.  We may now let $\Phi\to1$, obtaining
$$   |(f^t\ell)(A\circ\pi_U)-(f_X^t\ell_X)(A\circ\pi_{UX})|\le4\epsilon   $$
as announced.\qed
\medskip
	Proposition 3.2 also holds for the thermodynamic limit associated with a sequence $X_n\to\infty$
\bigskip\noindent
{\bf 4 $\Gamma$-states and their time evolution.}
\medskip
	We introduce now a special set of probability measures.
\medskip
	{\bf 4.1 Definition} ($\Gamma$-states).
\medskip
	{\sl We say that the probability measure $\ell$ carried by $[L]$ is a $\Gamma$-state if there exist constants $\tilde\beta_x>0$ (for $x\in L$), smooth functions $\tilde V_x:{\bf T}\to{\bf R}$ (for $x\in L$) and $\tilde W_{\{x,y\}}:{\bf T}\times{\bf T}\to{\bf R}$ (for $\{x,y\}\in\Gamma$) such that the $\tilde\beta_x$, $\tilde\beta_x^{-1}$, $\tilde V_x$, $\tilde W_{\{x,y\}}$, $\partial_{q_x}\tilde W_{\{x,y\}}$, $\partial_{q_y}\tilde W_{\{x,y\}}$ are bounded uniformly in $x,y\in L$, and the following holds:
\medskip
	For every finite $X\subset L$, the conditional measure $\ell_X(d\xi|\eta)$ of $\ell$ on $[X]$ given $\eta\in[L\backslash X]$ is of the form
$$	\ell_X(d\xi|\eta)=\hbox{{\rm const.}}
	\exp[-\sum_{x\in X}({1\over2}\tilde\beta_xp_x^2+\tilde V_x(q_x))
	-{\sum_{\{x,y\}}}^*\tilde W_{\{x,y\}}(q_x,q_y)]\,d\xi      $$
where $\sum^*$ extends over those ${\{x,y\}}\in\Gamma$ such that $x\in X$, and we have written $\xi=(p_x,q_x)_{x\in X}$, $\eta=(p_x,q_x)_{x\in L\backslash X}$.}
\medskip\noindent
[The $\Gamma$-states are {\it Gibbs states}\footnote{*}{See [14] for a discussion of Gibbs states in the simpler case of spin systems.  We shall not make use of the theory of Gibbs states in the present paper.} for a certain interaction given by the $\tilde\beta_x$, $\tilde V_x$, $\tilde W_{\{x,y\}}$].
\medskip
	If $\ell$ is a $\Gamma$-state we may, for finite $U\subset L$, write $(\pi_U\ell)(d\xi)=\ell_U(\xi)d\xi$ where $\ell_U$ is smooth on $[U]$.  Note that $\ell_U(\xi)$ has a $(p,q)$-factorization: it is the product of a smooth function of the $q_x$ for $x\in U$, and of a Gaussian $\sqrt{\tilde\beta_x/2\pi}\exp(-\tilde\beta_xp_x^2/2)$ for each $x\in U$.  
\medskip
	We shall now take $X$ finite and $Y=X\cup\tilde X_1\cup\ldots\cup\tilde X_{\bar k}$, where
$$	\tilde X_k=\{y\in L\backslash X:d(y,X)=k\}      $$
($\bar k$ is thus the ``size'' of $Y$).  If $\xi\in[X]$, and $\eta_k\in[\tilde X_k]$ for $k=1,\dots,\bar k$, the $\Gamma$-state property of $\ell$ then gives$$	\ell_Y(\xi,\eta_1,\ldots,\eta_{\bar k})
	=C_{\bar k}.\ell_0(\xi|\eta_1).\ell_1(\eta_1|\eta_2)\cdots
\ell_{\bar k-1}(\eta_{\bar k-1},\eta_{\bar k}).\ell_{\bar k}(\eta_{\bar k})  $$
where $C_{\bar k}$ is a normalization constant and (putting $\tilde W_{\{x,y\}}=0$ if $\{x,y\}\notin\Gamma$):
\medskip\noindent
$\displaystyle\ell_0(\xi|\eta_1)
	=\exp[-\sum_{x\in X}(\tilde\beta_xp_x^2/2+\tilde V_x(q_x))
	-\sum_{x,y\in X}\tilde W_{\{x,y\}}(q_x,q_y)
	-\sum_{x\in X}\sum_{y\in\tilde X_1}\tilde W_{\{x,y\}}(q_x,q_y)]$
\medskip\noindent
$\displaystyle\ell_k(\eta_k|\eta_{k+1})
	=\exp[-\sum_{x\in\tilde X_k}(\tilde\beta_xp_x^2/2+\tilde V_x(q_x))
	-\sum_{x,y\in\tilde X_k}\tilde W_{\{x,y\}}(q_x,q_y)$

$\displaystyle-\sum_{x\in\tilde X_k}\sum_{y\in\tilde X_{k+1}}\tilde W_{\{x,y\}}(q_x,q_y)]\qquad\hbox{when }k>0$, and 
\medskip\noindent
$\displaystyle\ell_{\bar k}(\eta_{\bar k})
=\int\nu(d\eta_{\bar k+1})\,\ell_{\bar k}(\eta_{\bar k}|\eta_{\bar k+1})\qquad\hbox{for some probability measure $\nu$ on $[\tilde X_{\bar k+1}]$}$.
\medskip
	Using the fact that the Jacobian of $f_U^t$ is $1$ ($f_U^t$ preserves $d\xi$) we have
$$	f_U^t((\pi_U\ell)(d\xi))=f_U^t(\ell_U(\xi)d\xi)
	=\ell_U(f_U^{-t}\xi)d\xi      $$
Thus, by Proposition 3.2, if $X\subset Y$ as above,
$$	(\pi_Xf^t\ell)(d\xi)
={w^*\lim}_{Y\to\infty}\pi_X(\ell_Y(f_Y^{-t}(\xi,\eta_Y))d\xi d\eta_Y)      $$
We may write 
$$	f_Y^{-t}(\xi,\eta_Y)=(f_{Y0}^{-t}(\xi,\eta_Y),
	f_{Y1}^{-t}(\xi,\eta_Y),\ldots,f_{Y\bar k}^{-t}(\xi,\eta_Y))      $$
with 
$$	f_{Y0}^{-t}(\xi,\eta_Y)\in[X],\qquad 
f_{Yk}^{-t}(\xi,\eta_Y)\in[\tilde X_k]\quad\hbox{for $k=1,\ldots,\bar k$}    $$
If $\xi,\tilde\xi\in[X]$, $\eta_Y\in[Y\backslash X]$, the quotient 
$$   {\ell_Y(f_Y^{-t}(\xi,\eta_Y))\over\ell_Y(f_Y^{-t}(\tilde\xi,\eta_Y))}   $$
is thus a product of quotients 
$$	{\ell_k(f_{Yk}^{-t}(\xi,\eta_Y)|f_{Y(k+1)}^{-t}(\xi,\eta_Y))\over
    \ell_k(f_{Yk}^{-t}(\tilde\xi,\eta_Y)|f_{Y(k+1)}^{-t}(\tilde\xi,\eta_Y))}
	\qquad\hbox{for $k=0,\ldots,\bar k-1$}      $$
and
$$	{\ell_{\bar k}(f_{Y\bar k}^{-t}(\xi,\eta_Y))\over
	\ell_{\bar k}(f_{Y\bar k}^{-t}(\tilde\xi,\eta_Y))}      $$
where the arguments $f_{Yk}^{-t}(\xi,\eta_Y)$, $f_{Y(k+1)}^{-t}(\xi,\eta_Y)$ and their derivatives have dependence on $\xi$ that decreases faster than exponentially with respect to $k$ (Propositions 2.7 and 2.8).
\medskip
	Let us define $\ell_{Yk}^t(\xi,\eta_Y)$ by
$$	\ell_k(f_{Yk}^{-t}(\xi,\eta_Y)|f_{Y(k+1)}^{-t}(\xi,\eta_Y))
=\ell_{Yk}^t(\xi,\eta_Y).\exp\sum_{x\in\tilde X_k}(-\tilde\beta_xp_x(0)^2/2) $$
for $k=0,\ldots,\bar k-1$, where $\tilde X_k$ is replaced by $X$ for $k=0$, and
$$	\ell_{\bar k}(f_{Y\bar k}^{-t}(\xi,\eta_Y))
	=\ell_{Y\bar k}^t(\xi,\eta_Y).
\exp\sum_{x\in\tilde X_{\bar k}}(-\tilde\beta_xp_x(0)^2/2)      $$
We shall also use $\ell_k^t(\xi,\eta)$ defined by
$$	\ell_k(f_k^{-t}(\xi,\eta)|f_{k+1}^{-t}(\xi,\eta))
=\ell_k^t(\xi,\eta).\exp\sum_{x\in\tilde X_k}(-\tilde\beta_xp_x(0)^2/2)      $$
where
$$  f^{-t}(\xi,\eta)=(f_0^{-t}(\xi,\eta),\ldots,f_k^{-t}(\xi,\eta),\ldots)  $$
with $f_0^{-t}(\xi,\eta)\in[X]$, and $f_k^{-t}(\xi,\eta)\in[\tilde X_k]$ for $k\ge1$.
\medskip
	From our definitions it follows that
$$	{\ell_Y(f_Y^{-t}(\xi,\eta_Y))\over\ell_Y(f_Y^{-t}(\tilde\xi,\eta_Y))}
	=\big[\prod_{k=0}^{\bar k}{\ell_{Yk}^t(\xi,\eta_Y)
        \over\ell_{Yk}^t(\tilde\xi,\eta_Y)}\big]
        \cdot{\exp\sum_{x\in X}(-\tilde\beta_xp_x(0)^2/2)
	\over\exp\sum_{x\in X}(-\tilde\beta_x\tilde p_x(0)^2/2)}  $$
\indent
	{\bf 4.2 Lemma} (basic uniform estimates).
\medskip
	{\sl In the above formula, we have, uniformly in $t\in[-T,T]$ and the size $\bar k$ of $Y$, the estimates
$$	|\log\ell_{Y0}^t(\xi,\eta_Y)|
	<\hbox{\rm const.}(1+\sup_{x\in X}|p_x(0)|)\leqno{(a)}      $$
$$    \big|\log{\ell_{Yk}^t(\xi,\eta_Y)\over\ell_{Yk}^t(\tilde\xi,\eta_Y)}\big|
	<{\hbox{\rm polyn.($k$)}\over k!}
(1+\sup_{x\in\tilde X_k}|p_x(0)|)\qquad\hbox{\rm if $k\ge1$}\leqno{(b)}     $$
These estimates remain true when $\ell_{Yk}^t$ is replaced by $\ell_k^t$.}
\medskip
	We note that, by Lemma 2.1(a),
$$	|p_x(t)^2-p_x(0)^2|\le K|t|(2|p_x(0)|+K|t|)      $$
From this, and the definitions, the first inequality of the lemma follows.  The second inequality is obtained by using also the finite dimensionality Assumption 1.1 and Lemma 2.1(b).\qed
\medskip
	Define now the regions $R_u,R^\times_v\subset[L]=[X]\times[L\backslash X]$ such that 
$$	R_u=\{(\xi,\eta):|p_x|\le u\hbox{ if }x\in X\}\qquad,\qquad R^\times_v
   =\{(\xi,\eta):|p_x|\le kv\hbox{ if }x\in\tilde X_k\hbox{ for }k\ge1\}   $$
\indent
	{\bf 4.3 Lemma} (existence of limit in $R_u\cap R_v^\times$).
\medskip
	{\sl In $R_u\cap R_v^\times$, the expression
$$	{\ell_Y(f_Y^{-t}(\xi,\eta_Y))
	\over\int_{[X]}d\tilde\xi\,\ell_Y(f_Y^{-t}(\tilde\xi,\eta_Y))}\cdot
[\ell_{Y0}^t(\xi,\eta_Y)\exp\sum_{x\in X}(-\tilde\beta_xp_x(0)^2/2)]^{-1}    $$
$$      =\Big[\int_{[X]}d\tilde\xi\,[\prod_{k=1}^{\bar k}
        {\ell_{Yk}^t(\tilde\xi,\eta_Y)\over\ell_{Yk}^t(\xi,\eta_Y)}]\,
        \ell_{Y0}^t(\tilde\xi,\eta_Y)
	\exp\sum_{x\in X}(-\tilde\beta_x\tilde p_x(0)^2/2)\Big]^{-1}      $$
has upper and lower bounds $\exp(\pm{\rm const.}(1+v))$ uniformly in $u$, $t\in[-T,T]$, and $\bar k$, and tends when $\bar k\to\infty$, uniformly for $(\xi,\eta)\in R_u\cap R^\times_v$, to
$$	\Big[\int_{[X]}d\tilde\xi\,[\prod_{k=1}^\infty{\ell_k^t(\tilde\xi,\eta)
	\over\ell_k^t(\xi,\eta)}]\,\ell_0^t(\tilde\xi,\eta)
	\exp\sum_{x\in X}(-\tilde\beta_x\tilde p_x(0)^2/2)\Big]^{-1}      $$
The limit is continuous.}  
\medskip
	Let $(\xi,\eta)\in R_u\cap R_v^\times$, and assume $\bar k$ to be large.  The quotients
$$   {\ell_{Yk}^t(\tilde\xi,\eta_Y)\over\ell_{Yk}^t(\xi,\eta_Y)}\qquad(k\ge1)   $$
are nearly independent of $Y$ ({\it i.e.}, of $\bar k$) for small $k$, and (using Lemma 4.2) very close to $1$ for large $k$, so that
$$	\lim_{\bar k\to\infty}
\prod_{k=1}^{\bar k}{\ell_{Yk}^t(\tilde\xi,\eta_Y)\over\ell_{Yk}^t(\xi,\eta_Y)}
=\prod_{k=1}^\infty{\ell_k^t(\tilde\xi,\eta)\over\ell_k^t(\xi,\eta)}      $$
uniformly, and we have bounds $\exp(\pm{\rm const.}(1+v))$ by Lemma 4.2(b).  Note now that $\ell_{Y0}^t(\tilde\xi,\eta_Y)$ tends to $\ell_0^t(\tilde\xi,\eta)$ uniformly for $(\tilde\xi,\eta)\in R_u\cap R_v^\times$, and we can extend the integral over $\tilde\xi$ from $|p_x|<u$ to $[X]$ because the Gaussian 
$$      \exp\sum_{x\in X}(-\tilde\beta_x\tilde p_x(0)^2/2)      $$ 
beats the exponential growth of $\ell_{Y0}^t(\tilde\xi,\eta_Y)$ given by Lemma 4.2(a).  Bounds of the form $\exp(\pm{\rm const.}(1+v))$ hold again after integration.\qed
\medskip
	{\bf 4.4 Lemma} (large $v$ Gaussian estimate).
\medskip
	{\sl For large $v$, $(\pi_{Y\backslash X}f_Y^t\pi_Y\ell)(d\eta_Y)$ has mass $<\exp(-\hbox{const.}v^2)$ outside of $\pi_{Y\backslash X}R^\times_v$, uniformly in the size $\bar k$ of $Y$.}
\medskip
	The $(p,q)$-factorization of $\ell_Y(\xi,\eta_Y)$ shows that the mass outside of $R^\times_v$ is bounded, uniformly in $\bar k$, by a Gaussian $<\exp(-\hbox{const.}v^2)$ for large $v$.  But the time evolution $f_Y^t$ changes $|p_x|$ (additively) by at most $K|t|$, so that the Gaussian estimate remains valid.\qed
\medskip
	{\bf 4.5 Theorem} (Smooth density of evolved states).
\medskip
	{\sl Let $\ell$ be a $\Gamma$-state.  For finite $X$, and $Y$ of size $\bar k$ as above, we write
$$	\bar\ell_{YX}^t(\xi)\exp\sum_{x\in X}(-\tilde\beta_xp_x(0)^2/2)
	=\int d\eta_Y\,\ell_Y(f_Y^{-t}(\xi,\eta_Y))      $$
There is a smooth function $\bar\ell_X^t(\xi)$ of $\xi$ and $t$ such that
$$	(\pi_Xf^t\ell)(d\xi)
	=\bar\ell_X^t(\xi)\exp\sum_{x\in X}(-\tilde\beta_xp_x(0)^2/2)\,d\xi      $$
and we have, uniformly for $|p_x|<u$ $(x\in X)$ and $|t|\le T$,
$$	\bar\ell_X^t(\xi)=\lim_{\bar k\to\infty}\bar\ell_{YX}^t(\xi)      $$
The limit also holds for the derivatives with respect to $\xi,t$.  The $\bar\ell_X^t(\xi)$, $\bar\ell_{YX}^t(\xi)$ have upper and lower bounds $\exp(\pm{\rm const.}(1+u))$, and the absolute values of their derivatives have bounds ${\rm polyn.}(u).\exp({\rm const.}(1+u))$ uniformly in $t\in[-T,T]$ and $\bar k$.}
\medskip
	We start with the remark that
$$	{\ell_Y(f_Y^{-t}(\xi,\eta_Y))\,d\xi\over
	\int_{[X]}d\tilde\xi\,\ell_Y(f_Y^{-t}(\tilde\xi,\eta_Y))}      $$
is the conditional measure of $f_Y^t\pi_Y\ell$ on $[X]$ given $\eta_Y\in[Y\backslash X]$.  Integrating this conditional measure with respect to $(\pi_{Y\backslash X}f_Y^t\pi_Y\ell)(d\eta_Y)$ yields $\pi_Xf_Y^t\pi_Y\ell$.  Thus
$$	\bar\ell_{YX}^t(\xi)=\int(\pi_{Y\backslash X}f_Y^t\pi_Y\ell)(d\eta_Y)
	{\ell_Y(f_Y^{-t}(\xi,\eta_Y))\over
	\int_{[X]}d\tilde\xi\,\ell_Y(f_Y^{-t}(\tilde\xi,\eta_Y))}
	\cdot\exp\sum_{x\in X}(\tilde\beta_xp_x(0)^2/2)      $$
The integrand in the right-hand side is the product of a factor controlled by Lemma 4.3, and a factor $\ell_{Y0}^t(\xi,\eta_Y)$ which has upper and lower bounds $\exp(\pm{\rm const.}(1+u))$ uniformly in $\bar k$ (by Lemma 4.2(a)) and tends to $\ell_0^t(\xi,\eta)$ when $\bar k\to\infty$, uniformly for $(\xi,\eta)\in R_u\cap R_v^\times$.  Using also Lemma 4.4 and the fact that $\pi_{Y\backslash X}f_Y^t\pi_Y\ell$ has the $w^*$limit $\pi_{L\backslash X}f^t\ell$ when $\bar k\to\infty$ (Proposition 3.2) we find that
$$	\lim_{\bar k\to\infty}\bar\ell_{YX}^t(\xi)      $$
$$	=\int(\pi_{L\backslash X}f^t\ell)(d\eta)\,\ell_0^t(\xi,\eta)
	\Big[\int_{[X]}d\tilde\xi\,[\prod_{k=1}^\infty{\ell_k^t(\tilde\xi,\eta)
	\over\ell_k^t(\xi,\eta)}]\,\ell_0^t(\tilde\xi,\eta)
	\exp\sum_{x\in X}(-\tilde\beta_x\tilde p_x(0)^2/2)\Big]^{-1}      $$
uniformly when $|p_x(0)|\le u$ for $x\in X$, with uniform upper and lower bounds $\exp(\pm{\rm const.}$ $(1+u))$.  We call the limit $\bar\ell_X^t(\xi)$.  Since
$$      \bar\ell_{YX}^t(\xi)\exp\sum_{x\in X}(-\tilde\beta_xp_x(0)^2/2)\,d\xi
        =(\pi_Xf_Y^t\pi_Y\ell)(d\xi)      $$
has the $w^*$ limit $(\pi_Xf^t\ell)(d\xi)$, it follows that this limit has a density 
$$      \bar\ell_X^t(\xi)\exp\sum_{x\in X}(-\tilde\beta_xp_x(0)^2/2)      $$
as asserted.
\medskip
	Using the notation 
$$    f^t(\hat\xi,\eta)=(f_0^t(\hat\xi,,\eta),f_{\hat\xi}^t(\eta))    $$
we may write
$$   \bar\ell_X^t(\xi)=\int(f^t\ell)(d\hat\xi\,d\eta)\,\Big[\int_{[X]}
d\tilde\xi\,[\prod_{k=1}^\infty{\ell_k^t(\tilde\xi,\eta)\over
\ell_k^t(\xi,\eta)}]\,l_0^t(\tilde\xi,\eta)\exp\sum_{x\in X}
(-\tilde\beta_x\tilde p_x(0)^2/2)\Big]^{-1}l_0^t(\xi,\eta)      $$
$$	=\int\ell(d\hat\xi\,d\eta)\,\Big[\int_{[X]}d\tilde\xi\,
	[\prod_{k=1}^\infty{\ell_k^t(\tilde\xi,f_{\hat\xi}^t(\eta))\over
	\ell_k^t(\xi,f_{\hat\xi}^t(\eta))}]l_0^t(\tilde\xi,f_{\hat\xi}^t(\eta))
	\exp\sum_{x\in X}(-\tilde\beta_x\tilde p_x(0)^2/2)\Big]^{-1}
	\,l_0^t(\xi,f_{\hat\xi}^t(\eta))      $$
and remember that this is the limit of a similar expression for $\bar\ell_{XY}^t(\xi)$.  We want to show that $\bar\ell_X^t(\xi)$ has derivatives (of all orders) with respect to $\xi,t$ by showing that the derivatives of $\bar\ell_{XY}^t(\xi)$, for $\xi,t$ in a compact set, are bounded with respect to $\bar k$.  Note that in estimating the integrals of polynomials in $p$, the $p_x(0)$ integral always has a Gaussian factor $\exp(-\beta_x p_x(0)^2/2)$ (remember the $(p,q)$ factorization of $\ell$).  Therefore we only have to worry about bounding the coefficients of the polynomials. 
\medskip
        Inspection of the above expression shows that computing a first order derivative essentially involves multiplying the integrand by the logarithmic derivatives of $l_0^t(\xi,f_{\hat\xi}^t(\eta))$, or $\ell_k^t(\tilde\xi,f_{\hat\xi}^t(\eta))/\ell_k^t(\xi,f_{\hat\xi}^t(\eta))$ and summing over $k$.  In view of the very explicit form of the logarithm of the $\ell_k$, we just have to estimate the derivatives of $f^{-t}(\xi,f_{\hat\xi}^t(\eta))$ with respect to $\xi$ and $t$.
\medskip
        As far as $\ell_0^t(\xi,f_{\hat\xi}^t(\eta))$ is concerned, we have to consider the derivatives of  $f_0^{-t}(\xi,f_{\hat\xi}^t(\eta))$ (or $f_1^{-t}(\xi,f_{\hat\xi}^t(\eta))$, which is similar).  The $\xi$-derivative is of the form $r_x^{(0,1)}(-t;x_1)$ with $x,x_1\in X$, giving a bounded contribution by Proposition 2.7.  The $t$-derivative is of the form
$$      r_x^{(1,0)}(-t)+\sum_{x_1}r_x^{(0,1)}(-t;x_1)r_{x_1}^{(1,0)}(t)      $$
with $x\in X$, and we may take $x_1\in\tilde X_{k_1}$.  By Proposition 2.7, $|r^{(0,1)}(\pm t)|$ is bounded by a polynomial in $p$ with bounded coefficients, and $r_x^{(0,1)}(-t;x_1)\le{\rm const.}/k_1!$, again giving a bounded contribution because $\sum_{k_1}|\tilde X_{k_1}|/k_1!$ is bounded.
\medskip
        We turn now to $\ell_k^t(\tilde\xi,f_{\hat\xi}^t(\eta))/\ell_k^t(\xi,f_{\hat\xi}^t(\eta))$, {\it i.e.}, we have to consider the derivatives of $f_k^{-t}(\xi,f_{\hat\xi}^t(\eta))$ (or $f_{k+1}^{-t}$).  The $\xi$-derivative is of the form $r_x^{(0,1)}(-t;x_1)$ with $x\in\tilde X_k,x_1\in X$, so that $|r_x^{(0,1)}(-t;x_1)|\le{\rm const.}/d(x,x_1)!={\rm const.}/k!$, which makes a bounded contribution because $\sum_k|\tilde X_k|/k!$ is bounded.  The $t$-derivative is of the form 
$$	\Delta r_x^{(1,0)}(-t;X)
        +\sum_{x_1}\Delta r_x^{(0,1)}(-t;x_1;X)r_{x_1}^{(1,0)}(t)      $$
where $x\in\tilde X_k$ and we may take $x_1\in\tilde X_{k_1}$.  By Proposition 2.8 we have
$$      |\Delta r_x^{(1,0)}(-t;X)|\le{\rm polyn.}(p)/k!      $$
$$      |\Delta r_x^{(0,1)}(-t;x_1;X)|
        \le{\rm const.}/\sigma!\le{\rm const.}/\max(k,k_1)!      $$
and since $\sum_k|\tilde X_k|/k!$, $\sum_{k,k_1}|\tilde X_k|.|\tilde X_{k_1}|/\max(k,k_1)!$ are bounded, we also have bounded contributions for the $t$-derivative.
\medskip
        We consider now higher order derivatives with respect to $\xi,t$.  The computation of such a derivative gives terms where the integrand is multiplied by a product of logarithmic derivatives of the type discussed above; the contribution is again seen to be bounded.  There are also terms containing derivatives of the logarithmic derivatives, and these are expressed in terms of higher order derivatives of $f^{-t}(\xi,f_{\hat\xi}^t(\eta))$ with respect to $\xi,t$.  The derivative $\partial^j/\partial\xi_{x_1}\cdots\partial\xi_{x_j}$ is estimated by $|r_x^{(0,j)}(-t;x_1,\ldots,x_j)|\le{\rm const.}/d(x,X)!$ giving a bounded contribution.  The derivative $\partial^i/\partial t^i$ is a sum of terms $\Delta r_x^{(i_1,i_2)}(-t;y_1,\ldots,y_{i_2};X)$ (multiplied by derivatives $\partial^kf_{\hat\xi}^t/\partial t^k=r^{(k,0)}$), where $i_1+i_2=i$; these terms can be estimated by Proposition 2.8, and give a bounded contribution.  The general mixed derivative $\partial^{i+j}/\partial t^i\partial\xi_{x_1}\cdots\partial\xi_{x_j}$, with $j\ge1$, is a sum of terms $r_x^{(i_1,j+i_2)}(-t;x_1,\ldots,x_j,y_1,\ldots,y_{i_2})$ with $x_1,\ldots,x_j\in X$ (multiplied by derivatives of the form $r^{(k,0)}$) which can be estimated by Proposition 2.7, and give a bounded contribution.\qed
\medskip
        {\bf 4.6 Remark} (uniform bounds).
\medskip
        The proof of Theorem 4.5 gives estimates of $\bar\ell_{YX}^t(\xi)$ and its derivatives with respect to $t$ and $\xi$, which are uniform with respect to the size $\bar k$ of $Y$.  They are also uniform with respect to the $\Gamma$-states $\ell$ with conditional measures corresponding to a fixed choice of $\tilde\beta_x$, $\tilde V_x$, $\tilde W_{\{x,y\}}$, and remain uniform if some of the $\tilde W_{\{x,y\}}$ are replaced by $0$.
\medskip
        For $Y$ finite and $\eta=(p_x,q_x)_{x\in Y}$ define 
$$      \tilde\ell_Y(\eta)=Z_Y^{-1}
        \exp[-\sum_{x\in Y}({1\over2}\tilde\beta_xp_x^2+\tilde V_x(q_x))
        -\sum_{x,y\in Y}\tilde W_{\{x,y\}}(q_x,q_y)]      $$
where $Z_Y^{-1}$ is a normalization factor, and write
$$      \bar{\tilde\ell}_{YX}^t(\xi)\exp\sum_{x\in X}(-\tilde\beta_xp_x(0)^2/2)
        =\int_{[Y\backslash X]}d\eta_Y\,\tilde\ell_Y(f_Y^{-t}(\xi,\eta_Y))      $$
Then the above remarks show that the uniform estimates on $\bar\ell_{YX}^t(\xi)$ and its derivatives given by Theorem 4.5 can be taken to hold also for $\bar{\tilde\ell}_{YX}^t$.
\bigskip\noindent
{\bf 5 Entropy.}
\medskip
        Given a $\Gamma$-state $\ell$, and $X\subset Y$ finite, we write
$$      \pi_X f_Y^t\pi_Y\ell=\ell_{YX}^t(\xi)\,d\xi
        \qquad,\qquad\pi_X f^t\ell=\ell_X^t(\xi)\,d\xi      $$
where $(t,\xi)\mapsto\ell_{YX}^t(\xi),\ell_X^t(\xi)$ are smooth on ${\bf R}\times[X]$.
\medskip
        In Theorem 4.5, we used the notation
$$      \ell_{YX}^t(\xi)
=\bar\ell_{YX}^t(\xi)\exp\sum_{x\in X}(-\tilde\beta_x p_x(0)^2/2)\qquad,\qquad
\ell_X^t(\xi)=\bar\ell_X^t(\xi)\exp\sum_{x\in X}(-\tilde\beta_x p_x(0)^2/2) $$
and saw that $\bar\ell_{YX}^t(\xi)$ tends to $\bar\ell_X^t(\xi)$ together with its derivatives, uniformly on compacts of ${\bf R}\times[X]$, when $Y\to\infty$.
\medskip
        We can now define a (Gibbs) entropy $S_Y^t(X)$ or $S^t(X)$ by
$$      S_Y^t(X)=-\int_{[X]}\ell_{YX}^t(\xi)\log\ell_{YX}^t(\xi)\,d\xi\qquad,
        \qquad S^t(X)=-\int_{[X]}\ell_X^t(\xi)\log\ell_X^t(\xi)\,d\xi      $$
These are convergent integrals in view of the uniform bounds given in Theorem 4.5.  Furthermore $S_Y^t(X)\to S^t(X)$, uniformly for $|t|\le T$, when $Y\to\infty$.
\medskip
        We may assume that $Y\supset\tilde X_1=\{y\in L:{\rm dist}(y,X)=1\}$.  If $\xi\in[X],\eta\in[Y\backslash X]$ or $[L\backslash X]$, let $\eta_1\in\tilde X_1$ be obtained from $\eta$ by restricting the index set to $\tilde X_1$.  Then the equations of motion for $\xi,\eta$ show that we may write
$$	{d\xi\over dt}={\cal X}(\xi,\eta_1)\qquad,\qquad
        {d\eta\over dt}={\cal Y}(\xi,\eta)      $$
where ${\cal X}$ does not depend on $Y$.  Writing $\hat\ell_Y^t=\ell_Y\circ f_Y^{-t}$, we have
$$     \ell_{YX}^t(\xi)=\int_{[Y\backslash X]}d\eta\,\hat\ell_Y^t(\xi,\eta)     $$
and the ``continuity equation''
$$      {d\over dt}\hat\ell_Y^t(\xi,\eta)
        +\nabla_\xi\cdot(\hat\ell_Y^t(\xi,\eta){\cal X}(\xi,\eta_1))
        +\nabla_\eta\cdot(\hat\ell_Y^t(\xi,\eta){\cal Y}(\xi,\eta))=0      $$
so that
$$      {d\over dt}\ell_{YX}^t(\xi)=-\int_{[Y\backslash X]}d\eta\,
        \nabla_\xi\cdot(\hat\ell_Y^t(\xi,\eta){\cal X}(\xi,\eta_1))
        =-\nabla_\xi\cdot\int_{[Y\backslash X]}d\eta\,
        \hat\ell_Y^t(\xi,\eta){\cal X}(\xi,\eta_1)      $$
$$      {d\over dt}\log\ell_{YX}^t(\xi)
        =-{1\over\ell_{XY}^t(\xi)}\nabla_\xi\cdot\int_{[Y\backslash X]}d\eta\,
        \hat\ell_Y^t(\xi,\eta){\cal X}(\xi,\eta_1)      $$
$$      {d\over dt}[\ell_{YX}^t(\xi)\log\ell_{YX}^t(\xi)]
        =-(\log\ell_{YX}^t(\xi)+1)\nabla_\xi\cdot\int_{[Y\backslash X]}d\eta\,
        \hat\ell_Y^t(\xi,\eta){\cal X}(\xi,\eta_1)      $$
\indent
        Using the estimates of Theorem 4.5 we find that $t\mapsto S_Y^t(X)$ is a smooth function of $t$, with
$$      {d\over dt}S_Y^t(X)=\int_{[X]}d\xi\,\log\ell_{YX}^t(\xi)\,\nabla_\xi
\cdot\int_{[Y\backslash X]}d\eta\,\hat\ell_Y^t(\xi,\eta){\cal X}(\xi,\eta_1)    $$
$$      =-\int_{[X]}{d\xi\over\ell_{YX}^t(\xi)}(\nabla_\xi\ell_{YX}^t(\xi))\cdot
  \int_{[Y\backslash X]}d\eta\,\hat\ell_Y^t(\xi,\eta){\cal X}(\xi,\eta_1)      $$
Write now $X+=X\cup\tilde X_1=\{y\in L:d(y,X)\le1\}$.  The probability measure 
$$      \ell_{YX+}^t(\xi,\eta_1)\,d\xi\,d\eta_1      $$
conditioned on $\xi\in[X]$ is denoted by $\ell_{YX+}^t(\eta_1|\xi)\,d\eta_1$ where 
$$      \ell_{YX+}^t(\eta_1|\xi)
        ={\ell_{YX+}^t(\xi,\eta_1)\over\ell_{YX}^t(\xi)}      $$
Theorem 4.5 gives uniform estimates for
$$      \ell_{YX+}^t(\eta_1|\xi)\qquad,
        \qquad\nabla_\xi\ell_{YX+}^t(\eta_1|\xi)      $$
so that
$$      {d\over dt}S_Y^t(X)=-\int_{[X]}d\xi\,\nabla_\xi\ell_{YX}^t(\xi)\cdot
\int_{[\tilde X_1]}d\eta_1\,\ell_{YX+}^t(\eta_1|\xi){\cal X}(\xi,\eta_1)     $$
$$      =\int_{[X]}d\xi\,\ell_{YX}^t(\xi)\nabla_\xi\cdot
\int_{[\tilde X_1]}d\eta_1\,\ell_{YX+}^t(\eta_1|\xi){\cal X}(\xi,\eta_1)     $$
It follows also that, when $Y\to\infty$, $dS_Y^t(X)/dt$ tends to
$$      \int_{[X]}d\xi\,\ell_X^t(\xi)\nabla_\xi\cdot
\int_{[\tilde X_1]}d\eta_1\,\ell_{X+}^t(\eta_1|\xi){\cal X}(\xi,\eta_1)      $$
uniformly for $|t|\le T$, and the limit is $dS^t(X)/dt$.
\medskip
	{\bf 5.1 Proposition} (time derivative of $S(X)$, $X$ finite).
\medskip
	{\sl When $Y\to\infty$, the derivative $dS_Y^t(X)/dt$ tends, uniformly for $|t|\le T$, to
$$	{d\over dt}S^t(X)=\int_{[X]}d\xi\,\ell_X^t(\xi)\nabla_\xi\cdot
\int_{[\tilde X_1]}d\eta_1\,\ell_{X+}^t(\eta_1|\xi){\cal X}(\xi,\eta_1)      $$
which is a smooth function of $t$.}
\medskip
        The proof, as given above, is essentially a corollary of Theorem 4.5.\qed
\medskip
	Suppose now that $\tilde X_1=\{y\in L:d(x,y)=1\}$ is finite, but $X$ is not necessarily finite.  We still have, for $Y$ finite,
$$	{d\over dt}S_Y^t(X\cap Y)=\int_{[X\cap Y]}d\xi\,\ell_{Y,X\cap Y}^t(\xi)\nabla_\xi\cdot
\int_{[\tilde X_1]}d\eta_1\,\ell_{Y,X+\cap Y}^t(\eta_1|\xi){\cal X}(\xi,\eta_1)      $$
and this can be bounded independently of $Y$.
\medskip
	{\bf 5.2 Proposition} (time derivative of $\Delta S^t(X)$).
\medskip
	{\sl If $X\subset L$, and $X$ is not necessarily finite, but $\tilde X_1=\{y\in L:d(X,y)=1\}$ is finite, we may define
$$	\Delta S^t(X)=\lim_{Y\to\infty}(S^t(X\cap Y)-S^0(X\cap Y))      $$
and we have
$$	{d\over dt}\Delta S_Y^t(X)=\int_{[X]}(\pi_Xf^t\ell)(d\xi)\nabla_\xi\cdot
\int_{[\tilde X_1]}d\eta_1\,\ell_{X+}^t(\eta_1|\xi){\cal X}(\xi,\eta_1)      $$
which is a smooth function of $t$.}
\medskip
This follows from the usual estimates.  Note that $\nabla_\xi$ is a derivative with respect to a finite number of variables corresponding to nonzero components of ${\cal X}(\xi,\eta_1)$.\qed
\medskip
	We shall now study a {\it conditional}, or "external" entropy $\check S$ defined for $X$ finite by
$$	\check S_Y^t(X)=S_Y^t(Y)-S_Y^t(Y\backslash X)      $$
Using the notation
$$	\hat\ell_Y^t=\ell_Y\circ f_Y^{-t}\qquad,\qquad
\ell_{Y,Y\backslash X}(\eta)=\int_{[X]}d\xi\,\hat\ell_Y^t(\xi,\eta)\qquad,\qquad
\ell_{YX}^t(\xi|\eta)={\hat\ell_Y^t(\xi,\eta)\over\ell_{Y,Y\backslash X}(\eta)}      $$
as above, we find
$$	\check S_Y^t(X)=-\int d\xi\,d\eta\,\hat\ell_Y^t(\xi,\eta)\log
{\hat\ell_Y^t(\xi,\eta)\over\ell_{Y,Y\backslash X}(\eta)}      $$
$$	=\int\ell_{Y,Y\backslash X}(\eta)\,d\eta[-\int d\xi\,\ell_{YX}^t(\xi|\eta)\log\ell_{YX}^t(\xi|\eta)]
	=\int\ell_{Y,Y\backslash X}(\eta)\,d\eta\,S_Y^t(X|\eta)      $$
where we have written
$$	\,S_Y^t(X|\eta)=-\int d\xi\,\ell_{YX}^t(\xi|\eta)\log\ell_{YX}^t(\xi|\eta)      $$
Let also
$$	\ell_X^t(\xi|\eta)=[\int_{[X]}d\tilde\xi[\prod_{k=1}^\infty{\ell_k^t(\tilde\xi,\eta)\over\ell_k^t(\xi,\eta)}]
\ell_0^t(\tilde\xi,\eta)\exp\sum_{x\in X}(-\tilde\beta_x\tilde p_x(0)^2/2)]^{-1}      $$
$$	\times\ell_0^t(\xi,\eta)\exp\sum_{x\in X}(-\beta_xp_x(0)^2/2)      $$
$$	S^t(X|\eta)=-\int d\xi\,\ell_X^t(\xi|\eta)\log\ell_X^t(\xi|\eta)      $$
then Lemma 4.2(a) and Lemma 4.3 imply that $S_Y^t(X|\eta)\to S^t(X|\eta)$ when $Y\to\infty$, uniformly for $\eta\in\pi_{L\backslash X}R_v^\times$ and $t\in[-T,T]$, and with uniform bounds
$$	|S_Y^t(X|\eta)|\le{\rm const.}(1+v)      $$
Therefore (using Lemma 4.4 and Proposition 3.2)we see that when $Y\to\infty$ we have
$$	\check S_Y^t(X)\to\check S^t(X)      $$
uniformly for $t\in[-T,T]$, where
$$	 \check S^t(X)=\int(\pi_{L\backslash X}f^t\ell)(d\eta)\,S^t(X|\eta)      $$
Note that $\check S^t(X)$ is obtained by taking the mean entropy $S^t(X|\eta)$ associated with $\ell_X^t(\xi|\eta)$ and averaging over $\eta$, while $S^t(X)$ is the entropy associated with the average $\ell_X(\xi)$ of $\ell_X^t(\xi|\eta)$ over $\eta$.  In particular, concavity gives $\check S^t(X)\le S^t(X)$.
\medskip
	Since $dS_Y^t(Y)/dt=0$, we have
$$	{d\over dt}\check S_Y^t(X)=-{d\over dt}S_Y^t(Y\backslash X)      $$
$$	=-\int_{[Y\backslash X]}d\eta\,\ell_{Y,Y\backslash X}^t(\eta)
\nabla_\eta\cdot\int_{[X]}d\xi\,\ell_{YX}^t(\xi|\eta){\cal Y}(\xi,\eta)      $$
$$	=-\int_{[Y\backslash X]}d\eta\,\ell_{Y,Y\backslash X}^t(\eta)\sum_{y\in\tilde X_1}
\partial_{\eta_y}\cdot\int_{[X]}d\xi\,\ell_{YX}^t(\xi|\eta){\cal Y}_y(\xi,\eta)      $$
where $\tilde X_1=\{y\in Y:d(y,X)=1\}$ and ${\cal Y}_y$ is the $y$-component of ${\cal Y}$.  We may now let $Y\to\infty$, finding:
\medskip
	{\bf 5.3 Proposition} (time derivative of the entropy $\check S$).
\medskip
	{\sl When $Y\to\infty$, the derivative $d\check S_Y^t(X)/dt$ tends, uniformly for $t\in[-T,T]$, to
$$	{d\over dt}\check S^t(X)=-\int(\pi_{L\backslash X}f^t\ell)(d\eta)\sum_{y\in\tilde X_1}
\partial_{\eta_y}\cdot\int_{[X]}d\xi\,\ell_X^t(\xi|\eta){\cal Y}_y(\xi,\eta)      $$
which is a smooth function of $t$.}
\medskip
	This is again a corollary of Theorem 4.5.\qed
\medskip
	Note that
$$	{d\over dt}\check S^t(X)=-{d\over dt}\Delta S^t(L\backslash X)      $$
\medskip
        {\bf 5.4 Assumption} (bounded energy).
\medskip
        {\sl For every finite $X\subset L$ the kinetic energy is bounded independently of $t$:}
$$      \int d\xi\,\ell_X^t(\xi)p_X^2/2\le{\rm const.}(X)      $$
[it would be equivalent to assume a bound on the total energy $H_X$].
\medskip
We have the general inequality
$$   S^t(X)\le\int d\xi\,\ell_X^t(\xi)p_X^2/2+|X|\log\sqrt{2\pi}      $$
[this follows from the ``variational principle for the free energy'', and can be proved by using the concavity of the log:
$$      \int d\xi\,\ell_X^t(\xi)\log{e^{-p_X^2/2}\over\ell_X^t(\xi)}
        \le\log\int d\xi\,e^{-p_X^2/2}=|X|\log\sqrt{2\pi}\qquad\qquad]      $$
Therefore the bounded energy assumption gives a bound on the entropy:
$$      S^t(X)\le{\rm const.}(X)      $$
Similarly, we find 
$$      S^t(X|\eta)\le\int d\xi\,\ell_X^t(\xi|\eta)p_X^2/2+|X|\log\sqrt{2\pi}      $$
hence
$$      \check S^t(X)\le\int d\xi\,\ell_X^t(\xi)p_X^2/2+|X|\log\sqrt{2\pi}\le{\rm const.}(X)    $$
In particular we have $\check S^t(X)\le S^t(X)\le{\rm const.}(X)$.
\medskip
        {\bf 5.5 Definitions} (large volume limit).
\medskip
        We may take a sequence $(T_n)$ tending to $+\infty$ such that $\displaystyle{1\over T_n}\int_0^{T_n}dt\,f^t\ell$ has a limit $\rho$ in the vague topology of measures on $[\dot L]$:
$$	t_n\to\infty\qquad,\qquad
	{1\over T_n}\int_0^{T_n}dt\,f^t\ell\to\rho      $$
We call the probability measure $\rho$ a {\it nonequilibrium steady state} (NESS).  In view of Assumption 5.4, $\rho$ is carried by $[L]$.  Furthermore $\rho$ is invariant under $(f^t)$.
\medskip
	We can also (by going to a subsequence) assume that 
$$	{\Delta S^{T_n}(X)\over T_n}
	={1\over T_n}\int_0^{T_n}dt\,{d\over dt}\Delta S^t(X)\to\sigma(X)      $$
when $\tilde X_1=\{y\in L:d(X,y)=1\}$ is finite ($X$ need not be finite).  Note that $\sigma(X)$ might not be determined by $\rho$ and $X$.
\medskip
	For notational simplicity we shall write $T\to\infty$ instead of $T_n,n\to\infty$.
\medskip
        {\bf 5.6 Interpretation} (entropy production). [16]
\medskip
	As mentioned in the Introduction, Denis Evans and coworkers [16] have proposed to identify the mean entropy production rate in a finite region $X$ to 
$$	e(X)=-\sigma(X)=-\lim_{T\to\infty}{S^T(X)-S^0(X)\over T}      $$
According to Proposition 5.1 this is the mean rate of volume contaction in $[X]$, and $e(X)$ corresponds to the accepted definition of entropy production in the presence of a deterministic thermostat.
\medskip
	A related choice is 
$$	\check e(X)=\sigma(L\backslash X)
	=-\lim_{T\to\infty}{\check S^T(X)-\check S^0(X)\over T}      $$
This is the mean rate of volume expansion in $[L\backslash X]$, and corresponds to the rate of entropy growth due to $X$, as seen by the "external world" $L\backslash X$.
\medskip
	Since  $\check S^t(X)\le S^t(X)\le{\rm const.}(X)$, we have $0\le e(X)\le\check e(X)$.
\medskip
	We may also define mean entropy production rates associated with a finite partition ${\cal A}=(X_0,X_1,\ldots,x_n)$ of $L$ provided $X_0,X_1,\ldots,X_n$ have finite "boundaries" $\{y\in L:d(X_i,y)=1\}$.  We write
$$	e({\cal A})=\sum_{j=0}^n\sigma(X_j)\qquad,
	\qquad\check e({\cal A})=\sum_{j:X_j\,{\rm infinite}}\sigma(X_j)      $$
In particular, in Case II, for $X$ finite $\supset X_0$, we have
$$	\check e(X)=\check e((X,L\backslash X))
	\le\check e((X,L_1\backslash X,L_2\backslash X))      $$
\medskip
	We proceed now with some general inequalities satisfied by $\sigma$, $e$, and $\check e$.
\medskip
	{\bf 5.7 Basic inequalities.}
\medskip
        We have $e(\emptyset)=\check e(\emptyset)=0$ by definition, and remember that $0\le e(X)\le\check e(X)$.  The strong subadditivity of the entropy implies that, if $U$, $V$ have finite boundaries,       
$$      \sigma(U\cup V)-\sigma(U)-\sigma(V)+\sigma(U\cap V)\le0      $$
(we have used the fact that $S^0((U\cup V)\cap Y)-S^0(U\cap Y)-S^0(V\cap Y)+S^0(U\cap V\cap Y)$ is bounded independently of $Y$).  This implies the strong superadditivity of $e$, and subadditivity of $\check e$.  In particular
$$      e(U\cup V)\ge e(U)+e(V)\qquad{\rm if}\qquad U\cap V=\emptyset      $$
and
$$      \check e(U\cup V)\le\check e(U)+\check e(V)      $$
If $U\subset V$ we also have\footnote{*}{Note that, for $U\subset V$, we have $S_Y^T(Y\backslash U)\le S_Y^T(Y\backslash V)+S_Y^T(V\backslash U)$, hence $\check S_Y^T(V)\le\check S_Y^T(U)+S_Y^T(V\backslash U)$, hence $\check e(V)\ge\check e(U)+e(V\backslash U)$.}
$$	e(U)\le e(V)\qquad,\qquad\check e(U)\le\check e(V)      $$
{\it i.e.}, $e(X),\check e(X)$ are increasing functions of $X$.
\medskip
	We can extend the definition of $e(X),\check e(X)$ to infinite $X$:
$$	e(X)=\sup_{{\rm finite}\,U\subset X}e(U)\qquad,\qquad
	\check e(X)=\sup_{{\rm finite}\,U\subset X}\check e(U)      $$
In the situations of interest for us $\check e(L)$ will be finite and we may call this quantity the {\it total entropy production rate}.  Note that the entropy production rate $\check e(X)$ is not an additive function of $X$, but that its subadditivity amounts to some kind of locality.  Note also that if $e(X)>0$ we must have $S^T(X)\to-\infty$, in particular the $\ell_X^T$ cannot remain bounded when $T\to\infty$, contrary to some evidence [4], [5].  But there is no obvious objection to having an entropy production rate $\check e(X)>0$.
\bigskip\noindent
{\bf 6 Thermodynamic bound on entropy production.}
\medskip
        We shall show that in Case I (an external force and a thermostat at temperature $\beta^{-1}$) we have
$$      \check e(X)\le\beta\times\hbox{energy flux to thermostat }      $$
where the right-hand side is the thermodynamic rate of entropy production.  A more general result is given below (see Proposition 6.3).
\medskip
        In Case I we have introduced a finite set $X_0$ on which external forces act.  As initial state $\ell$  we shall use the thermodynamic limit of a sequence:
$$	\ell={\theta\lim}_{Y\to\infty}\tilde\ell_Y(\eta)\,d\eta      $$
where\footnote{*}{More generally we could allow a term $\sum_{x\in X_0,y\notin X_0}\tilde W_{\{x,y\}}$ of interaction between $X_0$ and $Y\backslash X_0$.}
$$      \tilde\ell_Y(\eta)
        =Z_Y^{-1}\exp[-\tilde H_{X_0}-\beta H_{Y\backslash X_0}]      $$
In this formula, 
$$      \tilde H_{X_0}=\sum_{x\in X_0}({1\over2}\,\tilde\beta_xp_x^2+\tilde V_x)
        +\sum_{x,y\in X_0}\tilde W_{\{x,y\}}      $$
$$	H_{Y\backslash X}=\sum_{x\in Y\backslash X}({1\over2}\,p_x^2+V_x)
        +\sum_{x,y\in Y\backslash X} W_{\{x,y\}}      $$
and $Z_Y^{-1}$ is a normalization factor.  In this section it will be convenient to use $X_0$ instead of $X$ in the definition of $Y$, so that $Y=X_0\cup\tilde X_1\cup\ldots\cup\tilde X_{\bar k}$.  We take $X$ of the form $X_0\cup\tilde X_1\cup\ldots\cup\tilde X_k$ (this is no serious restriction) and choose a subsequence $\bar k\to\infty$ such that the $\pi_{XY}(\tilde\ell_Y(\eta)\, d\eta)$ converge vaguely (we use here the thermodynamic limit for a sequence as explained in Section 3).
\medskip
        The state $\ell$ is a $\Gamma$-state corresponding to the choice $\tilde\beta_x=\beta$, $\tilde V_x=\beta V_x$ for $x\notin X_0$, and $\tilde W_{\{x,y\}}=\beta W_{\{x,y\}}$ for $x,y\notin X_0$.  We define
$$	\tilde\ell_{YX}^t(\xi)
        =\int_{[Y\backslash X]}\tilde\ell_Y(f_Y^{-t}(\xi,\eta_Y))\,d\eta_Y    $$
and
$$      \tilde S_Y^t(X)
=-\int_{[X]}\tilde\ell_{YX}^t(\xi)\log\tilde\ell_{YX}^t(\xi)\,d\xi\qquad,\qquad 
\check{\tilde S}_Y^t(X)=\tilde S_Y^t(Y)-\tilde S_Y^t(Y\backslash X)      $$
Writing
$$      \tilde\ell_{YX}^t(\xi|\eta)
=\tilde\ell_{YY}^t(\xi,\eta)/\tilde\ell_{Y,Y\backslash X}(\eta)   $$
$$      \tilde S_Y^t(X|\eta)=-\int d\xi\,\tilde\ell_{YX}^t(\xi|\eta)\log\tilde\ell_{YX}^t(\xi|\eta)      $$
we find
$$      \check{\tilde S}_Y^t(X)=\int\tilde\ell_{Y,Y\backslash X}^t(\eta)d\eta\,\tilde S_Y^t(X|\eta)      $$
\indent
        {\bf 6.1 Lemma} (thermodynamic limit for $\check{\tilde S}$).
$$      \check{\tilde S}_Y^t(X)\to\check S^t(X)      $$
{\sl together with the $t$-derivatives, uniformly for $t\in[-T,T]$, when $\bar k\to\infty$}.
\medskip
        We have shown (in the proof of Proposition 5.2) how $\check S_Y^t(X)\to\check S^t(X)$.  We proceed in the same way here, using the uniform estimates of Theorem 4.5 which hold again when $\ell$ is replaced by $\tilde\ell$, as explained in Remark 4.6.\qed
\medskip
        Since $f_Y^t$ is volume preserving in $[Y]$ we have
$$	\tilde S_Y^t(Y)=\tilde S_Y^0(Y)      $$
therefore
$$    \check{\tilde S}_Y^0(X)-\check{\tilde S}_Y^t(X)
        =\tilde S_Y^t(Y\backslash X)-\tilde S_Y^0(Y\backslash X)      $$
We fix now $X$, with $X_0\subset X\subset Y$ as indicated above.  Note that, by the $(p,q)$ factorization, $\tilde S_Y^0(Y\backslash X_0)=\tilde S_{Y\backslash X_0}(Y\backslash X_0)$ is the sum of a momentum term $\tilde S^{0p}$ (integral over $p$, trivial) and a configuration term $\tilde S^{0q}$ (integral over $q$).  The configuration part $\tilde\ell_{Y,Y\backslash X}^q(q_{Y\backslash X})$ of $\tilde\ell_{Y,Y\backslash X}(q_{Y\backslash X})$ differs from $\tilde\ell_{Y,Y\backslash X_0}(q_Y)=\tilde\ell_{Y\backslash X_0}^q(q_{X\backslash X_0}(q_{Y\backslash X})$ by a factor bounded independently of $Y$ (because there is a finite number of bounded terms $\tilde V_x$ and $\tilde W_{\{x,y\}}$ with $x\in X\backslash X_0$).  Therefore $|\log\tilde\ell_{Y,Y\backslash X}^q(q_{Y\backslash X})-\log\tilde\ell_{Y\backslash X_0}^q(q_Y)|$ is bounded  independently of $Y$, hence $|\tilde S_Y^0(Y\backslash X_0)-\tilde S_Y^0(Y\backslash X)|\le C_0$ with $C_0$ independent of $Y$.  Define now a function $\ell^*$ on $[Y\backslash X_0]=[X\backslash X_0]\times[Y\backslash X]$ to be the product of 
$$      Z^{-1}\exp(-\beta p_{X\backslash X_0}^2/2)      $$ 
on $[X\backslash X_0]$ (with $Z^{-1}$ a normalization factor) and $\tilde\ell_{Y,Y\backslash X}^t$ on $[Y\backslash X]$.  Then there is a constant $C_1$ such that
$$	S^*=-\int_{[Y\backslash X_0]}d\eta\,\ell^*(\eta)
        \log\ell^*(\eta)=\tilde S_Y^t(Y\backslash X)+C_1      $$
and the ``variational principle for the free energy'' gives
$$	S^*-\tilde S_Y^0(Y\backslash X_0)\le\int_{[Y\backslash X_0]}d\eta\,
        (\ell^*(\eta)-\tilde\ell_{Y\backslash X_0}(\eta))\,
        \beta H_{Y\backslash X_0}(\eta)      $$
so that
$$    \tilde S_Y^t(Y\backslash X)-\tilde S_Y^0(Y\backslash X)
        \le C_0-C_1+S^*-\tilde S_Y^0(Y\backslash X)      $$
$$	\le C_0-C_1+\int_{[Y\backslash X_0]}d\eta\,
        (\ell^*(\eta)-\tilde\ell_{Y\backslash X_0}(\eta))\,
        \beta H_{Y\backslash X_0}(\eta)      $$
There are also constants $C_2,C_3$ such that
$$	\int_{[Y\backslash X_0]}d\eta\,\ell^*(\eta)\beta H_{Y\backslash X_0}(\eta)\le\int_{[Y\backslash X]}d\eta\,\tilde\ell_{Y,Y\backslash X}^t(\eta)\beta H_{Y\backslash X}(\eta)+C_2       $$
$$	\int_{[Y\backslash X_0]}d\eta\,\tilde\ell_{Y\backslash X_0}(\eta))\,\beta H_{Y\backslash X_0}(\eta)\ge\int_{[Y\backslash X]}d\eta\,\tilde\ell_{Y,Y\backslash X}^0(\eta)\beta H_{Y\backslash X}(\eta)+C_3      $$
Therefore, with a constant $C=C_0-C_1+C_2-C_3$ independent of $Y$ and $t$, we have
$$	\check{\tilde S}_Y^0(X)-\check{\tilde S}_Y^t(X)\le C+\int_{[Y\backslash X]}d\eta
[\tilde\ell_{Y,Y\backslash X}^t(\eta)-\tilde\ell_{Y,Y\backslash X}^0(\eta)]
	\beta H_{Y\backslash X}(\eta)      $$
$$	=C+\beta\int_{[Y]}d\eta\,[\tilde\ell_Y^t(\eta)-\tilde\ell_Y^0(\eta)]
	H_{Y\backslash X}(\eta)      $$
$$	=C+\beta\int_{[Y]}d\eta\,\tilde\ell_Y^0(\eta)
	[H_{Y\backslash X}(f_Y^t\eta)-H_{Y\backslash X}(\eta)]      $$
$$	=C+\beta\int_{[Y]}d\eta\,\tilde\ell_Y^0(\eta)
	\int_0^td\tau\,{d\over d\tau}H_{Y\backslash X}(f_Y^\tau\eta)      $$
The equations of motion yield
$$      {d\over d\tau}H_{Y\backslash X}(f_Y^\tau\eta)
        =\Phi(\pi_{X+,Y}f_Y^\tau\eta)      $$
where $X+=\{x\in L:d(x,X)\le1\}$ and $\Phi(p_Y,q_Y)$ is given by
$$      \Phi=-\sum_{x\in X}
\sum_{y\notin X}p_y{\partial\over\partial q_y}W_{\{x,y\}}(q_x,q_y)      $$
Therefore
$$      \int_{[Y]}d\eta\,\tilde\ell_Y^0(\eta)
\int_0^td\tau\,\Phi(\pi_{X+,Y}f_Y^\tau\eta)=\int_0^td\tau\,
\int_{[Y]}d\eta\,\tilde\ell_Y^\tau(\eta)\Phi(\pi_{X+,Y}\eta)      $$
$$   =\int_0^td\tau\,\int_{[X+]}d\xi\,\tilde\ell_{YX+}^\tau(\xi)\Phi(\xi)   $$
so that
$$	\check{\tilde S}_Y^0(X)-\check{\tilde S}_Y^t(X)
\le C+\beta\int_0^td\tau\int_{[X+]}d\xi\,\tilde\ell_{YX+}^\tau(\xi)\Phi(\xi) $$
We may now let $Y\to\infty$, obtaining
$$      \check S^0(X)-\check S^t(X)
\le C+\beta\int_0^td\tau\int_{[X+]}d\xi\,\ell_{X+}^\tau(\xi)\Phi(\xi)      $$
\indent
        {\bf 6.2 Proposition} (bound on entropy production, Case I).
\medskip
        {\sl In case I the mean rate of entropy production of the finite set $X$ is $\le\beta\times$energy flux out of $X_0$:
$$	0\le\check e(X)\le\beta\int(\pi_{X_0+}\rho)(d\xi)\Phi_0(\xi)      $$
where $\Phi_0$ is the function $\Phi$ computed for $X=X_0$.}
\medskip
        It suffices to consider the case of large $X$, so we assume $X\supset X_0$.  Taking in the previous inequality the large time limit described in 5.4 we obtain
$$	0\le\check e(X)\le\beta\int(\pi_{X+}\rho)(d\xi)\,\Phi(x)      $$
where the right-hand side is independent of $X$, and we may thus take $X=X_0$.\qed
\medskip
	We now give without proof a general bound on $\sigma(X)$, which can be obtained using the same ideas as for Proposition 6.2.
\medskip
	{\bf 6.3 Proposition.} (bound on $\sigma(X)$).
\medskip
	{\sl Let $X$ be infinite, with finite "boundary" $\tilde X_1=\{y\in L:d(X,y)=1\}$.  we let the initial state $\ell$ be the thermodynamic limit of a sequence:
$$	\ell={\theta\lim}_{Y\to\infty}\tilde\ell_Y(\eta)\,d\eta      $$
where
$$	\tilde\ell_Y(\eta)=Z_Y^{-1}e^{-\tilde H_Y}      $$
$$	\tilde H_Y=\sum_{x\in Y}({1\over2}\tilde\beta_xp_x^2+\tilde V_x)
	+\sum_{x,y\in Y}\tilde W_{\{x,y\}}      $$
We assume that 
$$	\tilde\beta_x=\beta\qquad,\qquad\tilde V_x=\beta V_x\qquad,
	\qquad\tilde W_{\{x,y\}}=\beta W_{\{x,y\}}      $$
when $x,y\in X$, {\it i.e.}, $X$ is a thermostat at temperature $\beta^{-1}$.  Then}
$$	\sigma(X)\le\beta\times\hbox{energy flux to $X$}      $$
\medskip
	Note that if the energy flows out of $X$, then $\sigma(X)<0$, and in particular $\sigma(X)$ does not vanish.  Applications of Proposition 6.3, in particular to Case II, are left to the reader.
\vfill\eject
\bigskip\noindent
{\bf A Appendices.}
\medskip
	{\bf A.1} (proof of Proposition 2.7).
\medskip
        By uniformity of the bounds it suffices to consider the situation when $Y$ is finite.  The case $(i,j)=(0,0)$ follows from Lemma 2.1(a), with $\tau=\sigma=0$.
\medskip
        Differentiating the time evolution equation for $\eta_x(t)=(p_x(t),q_x(t))$, we find
$$   {d\over dt}r_x^{(0,1)}=\sum_{y\in B_x^1}\Phi_{xy}(q_x,q_y)r_y^{(0,1)}   $$
where $\Phi_{xy}$ depends smoothly on the $q$'s, is independent of the $p$'s, and there is a uniform bound
$$      \sum_y|\Phi_{xy}|\le\bar K      $$
Therefore, if $r(t)=\sup_{x,x_1\in Y}|r_x^{(0,1)}(t;x_1)|$, we have
$$      |{d\over dt}r(t)|\le\bar Kr(t)      $$
with $r(0)=1$, so that $|r_x^{(0,1)}(t;x_1)|\le r(t)\le e^{\bar K|t|}$.
\medskip
        With the notation $k=d(x,x_1)$ we claim that
$$      |r_x^{(0,1)}(t;x_1)|\le\{e^{\bar K|t|}\}_k      $$
where we have defined $\{e^u\}_k=e^u-\sum_{n=0}^{k-1}u^n/n!\le e^u\inf_{\ell\le k}u^\ell/\ell!$  For $k=0$ the claim has been proved above.  For $k>0$ we have by induction
$$      |{d\over dt}r_x^{(0,1)}(t;x_1)|\le\bar K\{e^{\bar K|t|}\}_{k-1}      $$
and our claim follows by integration, using $r_x^{(0,1)}(0,x_1)=0$.
\medskip
        Remember that $\sigma(x,x_1,\ldots,x_j)$ is the smallest length of a subgraph of $\Gamma$ connecting $x,x_1,\ldots,x_j$.  We claim that for $j\ge1$ there are constants $L_j>0$ and $M_j>0$ such that
$$      |r_x^{(0,j)}(t;x_1,\ldots,x_j)|\le M_je^{j\bar K|t|}
        \inf_{0\le\tau\le\sigma}{(L_j|t|)^\tau\over\tau!}\eqno{(1)}      $$
where $\sigma=\sigma(x,x_1,\ldots,x_j)$.
\medskip
        We have already proved $(1)$ for $j=1$, with $L_1=\bar K$, $M_1=1$.  For $j>0$ we have
$$      {d\over dt}r_x^{(0,j)}(t;x_1,\ldots,x_j)
        =\sum_{y\in B_x^1}\Phi_{xy}r_y^{(0,j)}(t;x_1,\ldots,x_j)+rest\eqno{(2)}      $$
The $rest$ is a sum, over $y\in B_x^1$ and ${\cal X}$, of products of factors $r_{z_n}^{(0,j_n)}(t;X_n)$ where $z_n$ is $x$ or $y$, ${\cal X}=(X_n)$ is a partition of $(x_1,\ldots,x_j)$ into $|{\cal X}|>1$ subsequences of length $j_n$, and each product has a coefficient which is a smooth function of $q_x,q_y$.  Thus
$$      |rest|\le C_j\max_{\cal X}\exp(\sum_n j_n\bar K|t|)
        \prod_n{(L_{j_n}|t|)^{\tau_n}\over\tau_n!}
        \le C_je^{j\bar K|t|}{(L'_j|t|)^{\tau'}\over\tau'!}      $$
where $L'_j=\max_{\cal X}\sum_nL_{j_n}$ and $\tau'$ must be of the form $\sum_n\tau_n$, {\it i.e.}, $0\le\tau'\le\sum_n\sigma(z_n,X_n)$ where each $z_n$ is either $x$ or $y$, and therefore 
$$      \sum_n\sigma(z_n,X_n)+1\ge\sigma(x,x_1,\ldots,x_j)=\sigma      $$
so that all values of $\tau'$ between $0$ and $[\sigma-1]_+$ are allowed.  We have thus
$$      |rest|\le C_je^{j\bar K|t|}
\inf_{0\le\tau'\le[\sigma-1]_+}{(L'_j|t|)^{\tau'}\over\tau'!}      $$
\indent
        We shall prove $(1)$ by induction on $j$, assuming now $j>1$.  First let us write 
$$	\sup_x|r_x^{(0,j)}(t;x_1,\ldots,x_j)|=r(t:x_1,\ldots,x_j)
	=e^{\bar K|t|}s(t)      $$
and let
$$      \sigma'=\sigma(x_1,\ldots,x_j)=\min_x\sigma(x,x_1,\ldots,x_j)      $$
Then, for $0\le\tau'\le[\sigma'-1]_+$ and $t\ge0$,
$$      {d\over dt}r(t;x_1,\ldots,x_j)
\le\bar Kr(t;x_1,\ldots,x_j)+C_je^{j\bar Kt}{(L'_jt)^{\tau'}\over\tau'!}     $$
or
$$   {d\over dt}s(t)\le C_je^{(j-1)\bar Kt}{(L'_jt)^{\tau'}\over\tau'!}   $$
Thus
$$      {ds\over dt}
\le{d\over dt}[e^{(j-1)\bar Kt}{(C_jt)(L'_jt)^{\tau'}\over(\tau'+1)!}]
\qquad{\rm for}\qquad 0\le\tau'\le\sigma'-1      $$
and also (taking $\tau'=0$)
$$      {ds\over dt}\le{d\over dt}{C_j\over(j-1)\bar K}e^{(j-1)\bar Kt}      $$
We shall take $L_j=L'_j+C_j$ and $M_j=1+C_j/L'_j+C_j[(j-1)\bar K]^{-1}$.  In particular, we have
$$      s(t)\le M_je^{(j-1)\bar Kt}{(L'_jt)^{\tau'+1}\over(\tau'+1)!}
\qquad{\rm for}\qquad 0\le\tau'\le\sigma'-1      $$
$$      s(t)\le M_je^{(j-1)\bar Kt}      $$
hence
$$      s(t)\le M_je^{(j-1)\bar Kt}
        \inf_{0\le\tau\le\sigma'}{(L'_jt)^\tau\over\tau!}      $$
and
$$      |r_x^{(0,j)}(t:x_1,\ldots,x_j)|\le M_je^{j\bar K|t|}
        \inf_{0\le\tau\le\sigma'}{(L'_j|t|)^\tau\over\tau!}      $$
\indent
      To prove $(1)$, it suffices to show that if
$$      \sigma(x,x_1,\ldots,x_j)\ge\sigma'+k      $$
then
$$      |r_x^{(0,j)}(t;x_1,\ldots,x_j)|\le M_je^{j\bar K|t|}
        \inf_{0\le\tau\le\sigma'+k}{(L_j|t|)^\tau\over\tau!}      $$
and we have just shown this for $k=0$.  We proceed now by induction on $k$ for $k>0$.  For $y\in B_x^1$ we have
$$   \sigma(y,x_1,\ldots,x_j)\ge\sigma(x,x_1,\ldots,x_j)-1\ge\sigma'+k-1   $$
Therefore, our induction asssumption and estimate of $|rest|$ give
$$      |{d\over dt}r_x^{(0,j)}(t;x_1,\ldots,x_j)|\le\bar KM_je^{j\bar K|t|}
        \inf_{0\le\tau'\le\sigma'+k-1}{(L_j|t|)^{\tau'}\over\tau'!}
        +C_je^{j\bar K|t|}
        \inf_{0\le\tau'\le\sigma'+k-1}{(L'_j|t|)^{\tau'}\over\tau'!}      $$
Since $L_j=L'_j+C_j\ge C_j+\bar K$, and $1\le M_j$, we may write for $t\ge0$,
$$      {d\over dt}|r_x^{(0,j)}(t;x_1,\ldots,x_j)|\le M_je^{j\bar K|t|}
        \inf_{0\le\tau'\le\sigma'+k-1}L_j{(L_jt)^{\tau'}\over\tau'!}      $$
hence
$$|r_x^{(0,j)}(t;x_1,\ldots,x_j)|\le M_je^{j\bar K|t|}{(L_jt)^\tau\over\tau!}$$
if $1\le\tau\le\sigma'+k$, but the above inequality also holds by the induction assumption when $\tau=0$, and this completes the proof of $(1)$.
\medskip
        We discuss now the case $i>0$.  We have an explicit expression for $r^{(1,j)}=dr^{(0,j)}/dt$, given by the evolution equation for $(p,q)$ if $j=0$, by $(2)$ if $j\ge1$.  We may differentiate repeatedly with respect to $t$, replacing the derivatives in the right-hand side by using either the evolution equation for $(p,q)$ or $(2)$.  We express thus $r^{(i,j)}$ as a polynomial in the $p_y$ and the $r_y^{(0,\ell)}$ with $\ell\le j$, with coefficients that are smooth functions of the $q_y$.  The indices $y$ of the $p_y,q_y$, and $r_y^{(0,\ell)}$ that occur satisfy $d(x,y)\le i$.  Furthermore, in any given term of the polynomial, the factors $r_{y_1}^{(0,\ell_1)}(X_1),\ldots,r_{y_m}^{(0,\ell_m)}(X_m)$ that occur (with $(X_1,\ldots,X_m)$ forming a partition of $(x_1,\ldots,x_j)$) satisfy
$$      \sigma(x,y_1,\ldots,y_m)\le i      $$
$$      \sigma(x,y_1,\ldots,y_m)+\sigma(y_1,X_1)+\ldots+\sigma(y_m,X_m)
        \ge\sigma(x,x_1,\ldots,x_j)=\sigma      $$
so that
$$      \sigma(y_1,X_1)+\ldots+\sigma(y_m,X_m)\ge\sigma-i      $$
Therefore, using $(1)$, we see that $r^{(i,j)}$ is a polynomial of degree $\le i$ in the $p_y$ such that $y\in B_x^i$, with coefficients bounded in absolute value by
$$      {\rm const.}e^{j\bar K|t|}
        \inf_{0\le\tau\le\sigma-i}{(L_j|t|)^\tau\over\tau!}      $$
if $i\le\sigma$, otherwise by
$$      {\rm const.}e^{j\bar K|t|}      $$
concluding the proof of the proposition.\qed
\medskip
        {\bf A.2} (proof of Proposition 2.8).
\medskip
        Note that the conditions on the coefficients of ${\cal Q}$ are of the same form as those on the coefficients of ${\cal P}$ in Proposition 2.7, but $\sigma$ has a new definition (and the $L_j$, $M_{ij}$ may have to be chosen larger than for Proposition 2.7).
\medskip
        Note also that Lemma 2.1(b) gives, for $d(x,X)>0$,
$$	|\Delta r_x^{(0,0)}(t;X)|\le{(\bar K|t|)^\tau\over\tau!}
	\qquad{\rm if}\qquad 1\le\tau\le d(x,X)      $$
Using also Lemma 2.1(a), this proves the Proposition in the case $(i,j)=(0,0)$ since here $\sigma=d(x,X)$.
\medskip
        We shall later use the fact that for the $q$-component we have actually, for $d(x,X)\ge0$,
$$	|\Delta q_x(t;X)|=|q_x(t)-\tilde q_x(t)|\le{(\bar K|t|)^\tau\over\tau!}
	\qquad{\rm if}\qquad 0\le\tau\le d(x,X)      $$
[this is because $q_x(t),\tilde q_x(t)\in{\bf T}$, so that $|q_x(t)-\tilde q_x(t)|\le1$].
\medskip
        In the study of $\Delta r_x^{(0,j)}$ for $j>0$ we do not impose the condition $d(x,X)>0$.  If $j>0$, we claim that there are constants $L_j>0$ and $M_j>0$ such that
$$      |\Delta r_x^{(0,j)}(t;x_1,\ldots,x_j;X)|\le M_je^{j\bar K|t|}
        \inf_{0\le\tau\le\sigma}{(L_j|t|)^\tau\over\tau!}\eqno{(3)}      $$
where $\sigma=\sigma(x,x_1,\ldots,x_j;X)$.
\medskip
        This will be proved by induction on $j$, using $(1)$ and the equation
$$	{d\over dt}\Delta r_x^{(0,j)}(t;x_1,\ldots,x_j;X)
=\sum_{y\in B_x^1}\Phi_{xy}(q_x,q_y)\Delta r_y^{(0,j)}(t;x_1,\ldots,x_j;X)
        +rest      $$
The $rest$ here is a finite sum of products, each of which has exactly one factor with a $\Delta$ in front of it.  The factors are: a coefficient depending smoothly on $q_x,q_y$, and factors $r_{z_n}^{(0,j_n)}$ where $z_n$ is $x$ or $y$ and the $X_n$ form a partition ${\cal X}=(X_n)$ of $(x_1,\ldots,x_j)$ into $|{\cal X}|$ subsequences of length $j_n$.
\medskip
        In particular, for $j=1$, the $rest$ is $\sum_y\Delta\Phi_{xy}.r_y^{(0,1)}$.  Using the remark above on $|\Delta q_x|$, and the earlier bound on $|r_x^{(0,1)}|$ one finds
$$	|rest|\le{\rm const.}{(\bar K|t|)^k\over k!}.
        e^{\bar K|t|}{(\bar K|t|)^\ell\over\ell!}
\le{\rm const.}e^{\bar K|t|}{(2\bar K|t|)^{k+\ell}\over(k+\ell)!}      $$
where $k$ is allowed values in $[0,d(y,X)]$ or $[0,d(x,X)]$ and $\ell$ is allowed values in $[0,d(y,x_1)]$, so that $k+\ell$ is allowed all values in $[0,[\sigma-1]_+]$.
\medskip
        For general $j>0$, using induction on $j$, and the bounds on $|\Delta q_x|,|r_x^{(0,j)}|$ shows that the products appearing in the $rest$ have, in absolute value, bounds of the form
$$      {\rm const.}{(\bar K|t|)^k\over k!}.(\exp\sum_nj_n\bar K|t|)
        .\prod_n{(L_{j_n}|t|)^{\ell_n}\over\ell_n!}      $$
$$      \le{\rm const.}e^{j\bar K|t|}
        {[(\bar K+\sum_nL_{j_n})|t|]^{k+\sum\ell_n}\over(k+\sum\ell_n)!}
        \le{\rm const.}e^{j\bar K|t|}{(L'_j|t|)^{\tau'}\over\tau'!}      $$
where $L'_j=\bar K+\max_{\cal X}\sum_nL_{j_n}$, and we must now discuss the range of $\tau'=k+\sum\ell_n$.  Remember that there is a $\Delta$ in front of one of the factors of the product we are considering.  
\medskip
        If the $\Delta$ is in front of the coefficient depending smoothly on $q_x,q_y$, this corresponds to $k\in[0,d(z,X)]$ with $z=x$ or $y$, while $\ell_n\in[0,\sigma(z_n,X_n)]$ with $z_n=x$ or $y$.  Since $d(x,y)=1$, we have $d(z,X)+\sum_\ell\sigma(z_\ell,X_\ell)+1\ge\sigma(x,x_1,\ldots,x_j;X)=\sigma$; therefore $\tau'=k+\sum\ell_n$ is allowed all values such that $0\le\tau'\le[\sigma-1]_+$.
\medskip
        If the $\Delta$ is in front of one of the $r_{z_n}^{(0,j)}$, say for $n=a$, the corresponding $\ell_a$ is $\in[0,\sigma(z_a,X_a;X)]$ by the induction assumption, the other $r_{z_n}^{(0,j_n)}$ are $\in[0,\sigma(z_n,X_n)]$, and we have $k=0$.  Note that
$$      \sigma(z_a,X_a;X)+\sum_{n\ne a}\sigma(z_n,X_n)+1
        \ge\sigma(x,x_1,\ldots,x_j;X)=\sigma      $$
therefore $\tau'=\sum\ell_n$ is again allowed all values such that $0\le\tau'\le[\sigma-1]_+$.  In conclusion we have
$$      |rest|\le C_je^{j\bar K|t|}
        \inf_{0\le\tau'\le[\sigma-1]_+}{(L'_j|t|)^{\tau'}\over\tau'!}      $$
\indent
        To start the proof of $(3)$ we write
$$	\sup_x|\Delta_x^{(0,j)}(t;x_1,\ldots,x_j;X)|=r(t;x_1,\ldots,x_j;X)
        =e^{\bar K|t|}s(t)      $$
and let $\sigma'=\sigma(x_1,\ldots,x_j;X)=\min_x\sigma(x,x_1,\ldots,x_j;X)$,  Then for $\tau'\in[0,[\sigma'-1]_+]$ and $t\ge0$ we obtain, as in the proof of Proposition 2.7
$$    {d\over dt}s(t)\le C_je^{(j-1)\bar Kt}{(L'_jt)^{\tau'}\over\tau'!}    $$
hence
$$      |\Delta r_x^{(0,j)}(t;x_1,\ldots,x_j;X)|
\le M_je^{j\bar K|t|}\inf_{0\le\tau\le\sigma'}{(L_j|t|)^\tau\over\tau!}      $$
and the proof of $(3)$ continues as the proof of $(1)$.
\medskip
      The case $i>0$ (taking now $d(x,X)>i$) is treated as in the proof of Proposition 2.7.\qed
\vfill\eject\noindent
{\bf References.}
\medskip
[1] L. Andrey.  ``The rate of entropy change in non-Hamiltonian systems.''  Phys. Letters {\bf 11A},45-46(1985).

[2] F. Bonetto, A. Kupiainen, and J.L. Lebowitz. ``Absolute continuity of projected SRB measures of coupled Arnold cat map lattices'' Ergod. Th. Dynam. Syst. {\bf 25},59-88(2005).

[3] O. Bratteli and D.W. Robinson.  {\it Operator algebras and quantum statistical mechanics} {\bf 2}, second edition.  Springer, Berlin, 1997.

[4] J.-P. Eckmann, C.-A. Pillet, and L. Rey-Bellet.  "Nonequilibrium statistical mechanics of anharmaonic chains coupled to two baths at different temperatures"  Commun. Math. Phys. {\bf 201},657-697(1999).

[5] J.-P. Eckmann, C.-A. Pillet, and L. Rey-Bellet.  "Entropy production in non-linear, thermally driven Hamiltonian systems"  J. Stat. Phys. {\bf 95},305-331(1999).

[6] D.J. Evans, E.G.D. Cohen, and G.P.Morriss.  ``Probability of second law violations in shearing steady states.''  Phys. Rev. Letters {\bf 71},2401-2404(1993).

[7] D.J. Evans and G.P. Morriss.  {\it Statistical mechanics of nonequilibrium fluids.}  Academic Press, New York, 1990.

[8] G. Gallavotti ``Stationary nonequilibrium statistical mechanics'' preprint arXiv: cond-mat/0510027 v1  2 Oct 2005.

[9] G. Gallavotti and E.G.D. Cohen.  ``Dynamical ensembles in stationary states.'' J. Statist. Phys. {\bf 80},931-970(1995).

[10] W.G. Hoover.  {\it Molecular dynamics.}  Lecture Notes in Physics {\bf 258}.  Springer, Heidelberg, 1986.

[11] V. Jak\v si\'c and C.-A. Pillet.  "On entropy production in quantum statistical mechanics"  Commun Math. Phys. {\bf217},285-293(2001).

[12] O.E. Lanford, J.L. Lebowitz, and E.H. Lieb. ``Time evolution of infinite anharmonic systems'' J. Statist. Phys. {\bf 16},453-461(1977).

[13] H.A. Posh, and W.G. Hoover.  "Large-system phase-space dimensionality loss in stationary heat flows"  Physica D{\bf 187},281-293(2004).

[14] D. Ruelle  {\it Thermodynamic Formalism.}  Addison-Wesley, Reading MA, 1978.

[15] D. Ruelle  ``Entropy production in quantum spin systems.''  Commun. Math. Phys. {\bf 224},3-16(2001).

[16] S.R. Williams, D.J. Searles, and D.J. Evans.  "Independence of the transient fluctuation theorem to thermostatting details" Phys. Rev. E {\bf 70},066113--1-6(2004).
\end